\documentclass{aa}  
\usepackage{float}
\usepackage[left]{lineno}
\usepackage{graphicx}
\usepackage{comment}
\usepackage{longtable}

\usepackage[colorlinks=true,linkcolor=purple,citecolor=blue,filecolor=black,urlcolor=cyan,final=true]{hyperref}

\usepackage{xcolor}
\usepackage{txfonts}

\newcommand{\divV}{\nabla \cdot \vec{V}}

\begin{document} 

   \title{
   {Freely propagating flanks of wide coronal-mass-ejection-driven shocks: Modelling and observational insights}
   }

   \titlerunning{Freely propagating flanks of wide CME-driven shocks
   }
   \subtitle{}

    \author{
        N.~Wijsen\inst{1}
        \and
        I.~C.~Jebaraj\inst{2}
        \and
        N.~Dresing\inst{2}
        \and
        A.~Kouloumvakos\inst{3}
        \and 
        E.~Palmerio\inst{4}
        \and
        L.~Rodr\'iguez-Garc\'ia\inst{5,6}
        }
        
   \institute{      
            Center for mathematical Plasma Astrophysics, KU Leuven, Kortrijk/Leuven, Belgium  
        \and
            Department of Physics and Astronomy, University of Turku, Turku, Finland  
        \and    
            The Johns Hopkins University Applied Physics Laboratory, Laurel, MD, USA
        \and  
            Predictive Science Inc., San Diego, CA, USA 
        \and 
            European Space Astronomy Centre, European Space Agency, Villanueva de la Ca{\~n}ada, Madrid, Spain 
        \and 
            Universidad de Alcalá, Space Research Group (SRG-UAH), Alcalá de Henares, Madrid, Spain 
        }
  
   \date{}

  \abstract
{
Widespread solar energetic particle (SEP) events remain poorly understood phenomena in space weather. These events are often linked to coronal mass ejections (CMEs) and their shocks, but the mechanisms governing their global particle distribution remain debated.
{The 13 March 2023 event is particularly notable as a widespread SEP event associated with an exceptionally fast interplanetary shock.\ With speeds of up to 3000 km/s, it is one of the most extreme shocks observed in recent years.}
}
{
{We aim to investigate whether {the flanks of a} wide CME-driven shock can decouple from the CME and continue propagating as freely propagating shock waves. If shocks are the primary SEP source, such a mechanism could help explain some of the widest SEP events.}
}
{
{Using EUHFORIA, a 3D magnetohydrodynamic heliospheric model, we simulated the evolution of wide CME-driven shocks. We modified the model to allow direct shock injection at the inner boundary, upstream of the CME ejecta. Applying this to the 13 March 2023 event, we modelled two  simultaneous CMEs whose shocks form a single, wide shock envelope that spans 280$^\circ$ in longitude. We then compared our results to in situ observations.}
}
{
{Our simulations show that the flanks of wide CME shocks can persist as freely propagating waves beyond 2~au.   
For the 13 March 2023 event, the modelled shock arrival times and amplitudes of associated plasma parameters (e.g. speed and density) show good agreement with observations from various spacecraft distributed across different radial distances and longitudes. 
Furthermore, the combined shock structure expands into a quasi-circumsolar wave as it propagates outwards. }
}
{ 
{These findings indicate that the shock flanks of fast CMEs can persist for a long time, supporting the idea that such freely propagating shock flanks play a key role in the global distribution of SEPs in widespread events.}
}

   \keywords{Shocks --
            Coronal Mass Ejections --- 
            Solar energetic particles
               }
   \maketitle
%

\section{Introduction}\label{sec:intro}

Widespread solar energetic particle (SEP) events, spanning more than 180\(^{\circ}\) around the Sun, remain poorly understood in heliophysics. They are often associated with large coronal mass ejections (CMEs) and the shocks they generate \citep{Rouillard2012, Lario2014}, but the exact mechanisms that enable particles to spread so widely are still debated \citep[e.g.][]{Kollhoff2021, 2021Rodriguez-Garcia, Kouloumvakos2022, Dresing2023}. One possibility involves wide, possibly circumsolar, shock waves \citep[e.g.][]{Hou2022}, which can accelerate particles throughout extensive regions of the heliosphere \citep{gomez-herrero15, lario16}. Observations of large shock extents \citep[e.g.][]{Nitta2013, Kwon2017} support this view. Alternatively, SEPs may originate from localized regions and spread via cross-field transport in the corona and interplanetary (IP) medium \citep[e.g.][]{dresing12, strauss17b}.

Here, we focus on the role of wide shocks in driving widespread SEP events. A key question is how far and for how long shocks — which are often significantly wider than their parent CMEs — can persist. Although the solar wind and corona are complex, {shocks typically evolve as they propagate, with some regions remaining CME-driven while others expand beyond direct CME influence \citep[e.g.][]{Corona-Romero2011,Warmuth2015}. This transition  begins at the shock flanks, which decouple first, while the central region remains CME-driven. Over time, as the CME slows and its ability to drive the shock diminishes, even the central part of the shock can detach, leaving behind a fully decoupled wavefront \citep[e.g.][]{Pinter1990}.}

Observations by \citet{Kwon2018} highlight that shock strength can remain substantial well beyond the region directly driven by the CME apex, indicating that wide CME-driven shocks can accelerate energetic particles over larger spatial and temporal scales than the CME width alone would suggest. Shock modelling and reconstructions generally support these observations \citep[e.g.][]{kouloumvakos19, Jebaraj21}, though there are exceptions \citep[e.g.][]{lario17, Kollhoff2021}, illustrating the complexity of shock evolution and SEP transport.

A related issue is the fate of shocks as they propagate into IP space. {Once partially or fully decoupled from their drivers, shocks rely on the energy and momentum they gained during their driving phase \citep[e.g.][]{Corona-Romero2012}. Their subsequent evolution is reminiscent of a blast wave \citep[e.g.][]{Smart1985,Pinter1990,Corona-Romero2011,Li2024ApJ} and depends strongly on the upstream conditions, such as the solar wind density, magnetic field structure, and speed. In some cases, shocks can maintain sufficient strength to accelerate particles over large distances and for a long time, while others weaken relatively quickly \citep{Liu17}.}

{A key factor in this evolution is the decrease in the local fast magnetosonic speed with heliocentric distance. For instance, \citet{Echer2019} show that IP shocks near Jupiter’s orbit can exhibit higher Mach numbers on average than those at 1\,au, not because of additional energy from the CME, but because the magnetosonic speed at 5~au is only about 60\% of its value at 1~au. Consequently, shocks that appear modest closer to the Sun can seem stronger farther out simply due to the changing local solar wind environment. Understanding these factors is essential for accurately modelling how long shocks can persist and the conditions under which they can continue to accelerate SEPs throughout the heliosphere.}

{In this work we explored the possibility that the 13 March 2023 event involved two nearly simultaneous CMEs whose combined eruption drove a wide IP shock. We examine whether the flanks of this shock could have decoupled from the ejecta and evolved into freely propagating waves as the structure expanded through the heliosphere. To investigate this scenario, we used the European Heliospheric Forecasting Information Asset \citep[EUHFORIA;][]{pomoell18}, which solves 3D, time-dependent ideal magnetohydrodynamic (MHD) equations from 0.1~au onwards. As part of this study, we upgraded EUHFORIA to enable the direct injection of shocks, rather than relying solely on ejecta-driven shock waves (see Appendix~\ref{app:EUHFORIA}). 
In our simulations, a magnetic ejecta continues to drive the shock at its apex, while the shock flanks can propagate freely. 
Under these modelling assumptions, the shock can maintain its strength out to large radial distances, partly due to the decreasing fast magnetosonic speed farther from the Sun. Our results show promising agreement with in situ observations at various longitudes and radial distances, suggesting that such wide shocks can persist well beyond the CME core.}

\begin{figure}
    \centering
    \includegraphics[width=0.99\linewidth]{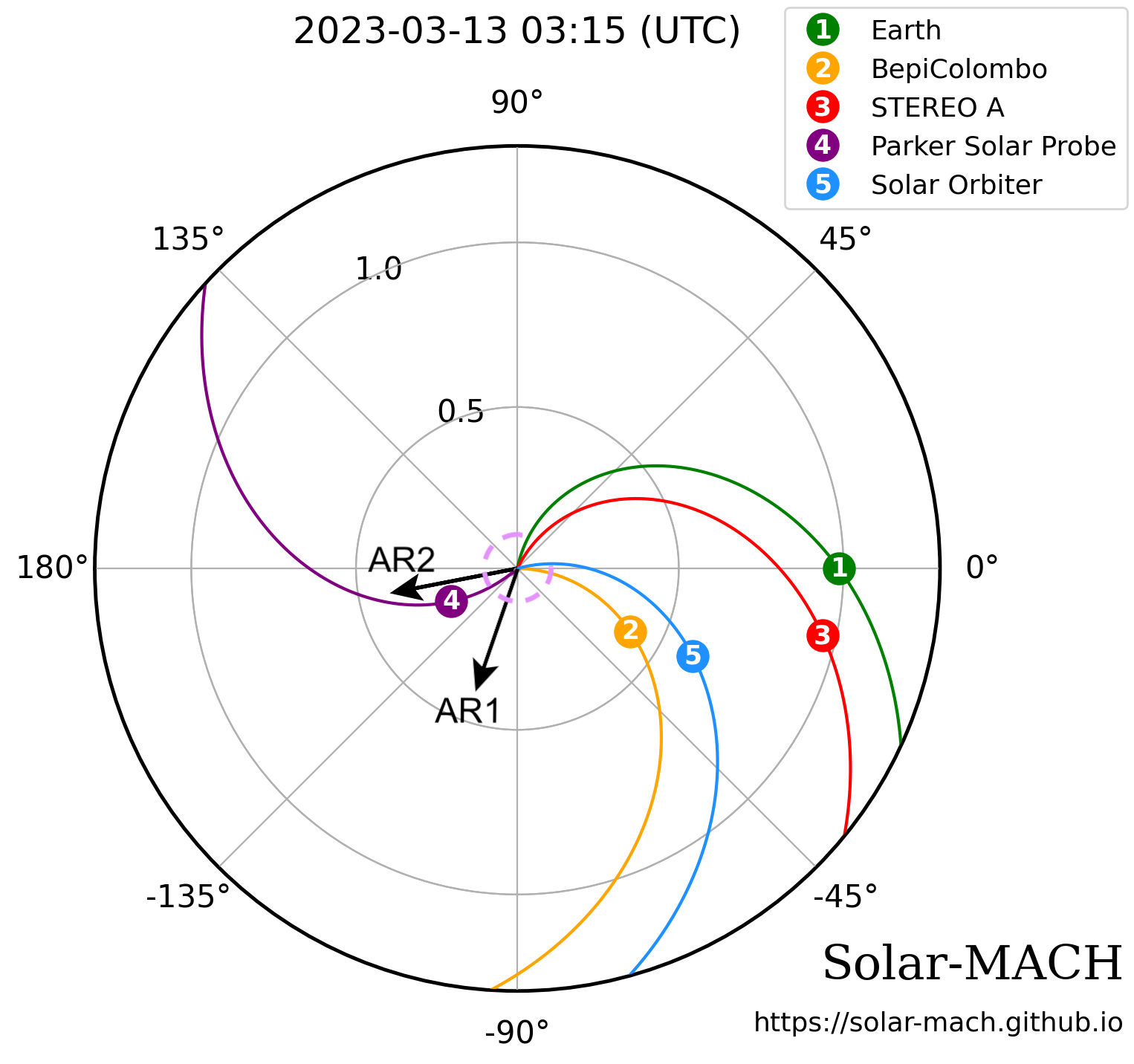}
    \caption{Spacecraft constellation in the solar equatorial plane, showing nominal Parker spiral field lines connecting each spacecraft to the Sun. The field lines were derived using measured solar wind speeds for STEREO-A, Earth, and Parker Solar Probe, while for Solar Orbiter and BepiColombo, where no solar wind plasma data were available at the time of the event, solar wind speeds from the EUHFORIA simulation are used. The two arrows indicate the estimated locations of the ARs from which the CMEs erupted. The dashed pink line represents the combined extent of the two shock waves injected in the EUHFORIA simulation. This figure was made using the Solar-Mach tool \citep{giesler2023}.}
    \label{fig:sc_constellation}
\end{figure}
\begin{figure*}
    \centering
    \includegraphics[width=0.9\linewidth]{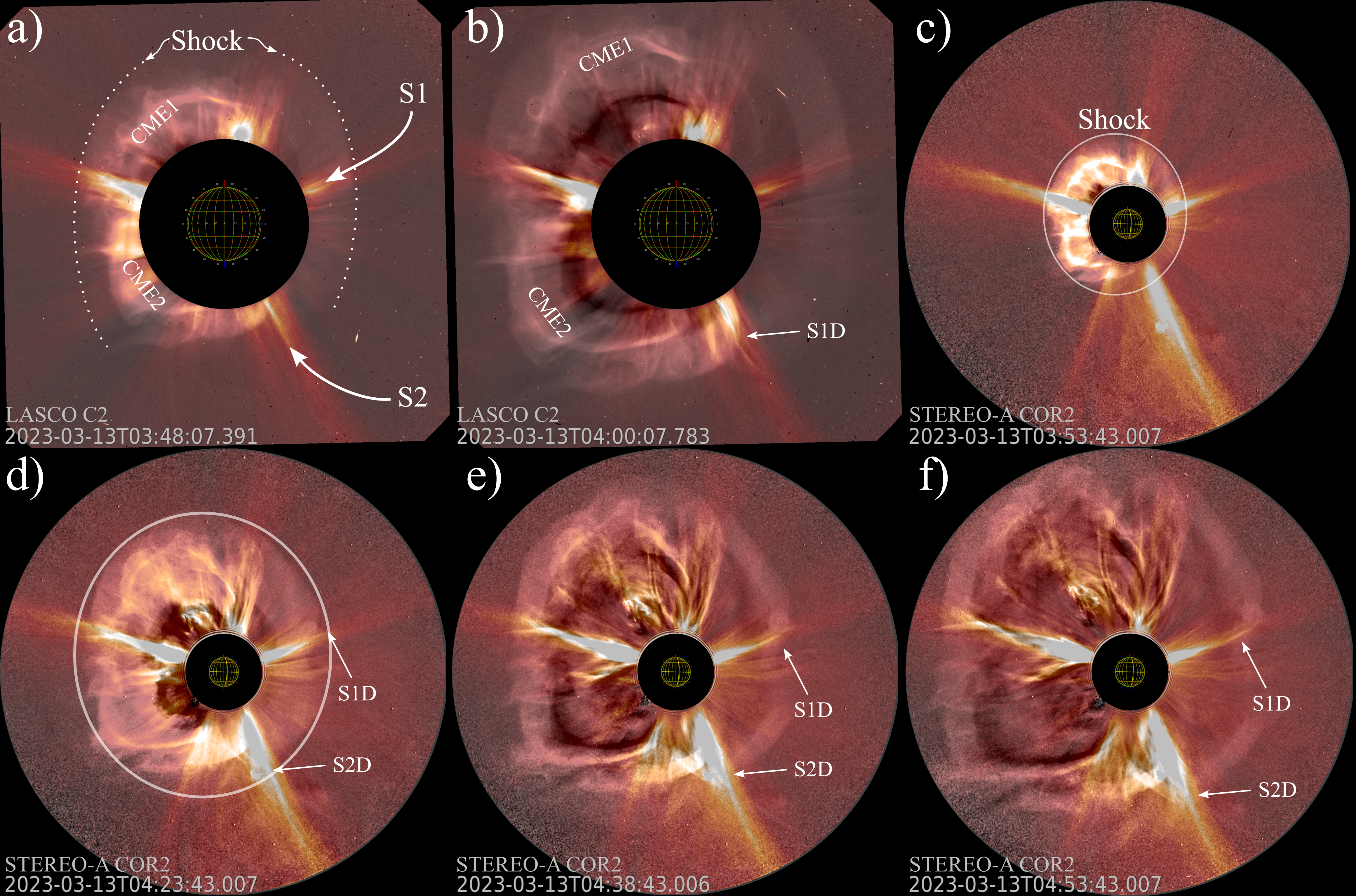}
    \caption{Selected snapshots of the CME evolution on 13 March 2023, as seen in coronagraphic observations from LASCO and STEREO-A. Each panel shows an overlay of a plain coronagraphic image and the corresponding running difference image at the same time from LASCO-C2 (panels a and b) and COR2-A (panels c to f). At selected frames we note the CMEs and outline the location of the CME-driven shock wave. The interaction of the shock with streamers (S1 and S2) are also noted in panels d to f (streamers deflections S1D and S2D). The various features are discussed in detail in Sect.~\ref{sec:obs}.}
    \label{fig:coronagraph}
\end{figure*}

\section{The 13 March 2023 event: Observations}\label{sec:obs}

On 13 March 2023 at approximately 03:15~UT,  significant solar activity on the Sun’s far side produced a fast and wide shock wave. Parker Solar Probe \citep[][]{fox16}, positioned closest to the eruption site, observed an extraordinary shock wave followed by a magnetic cloud at 0.24~au. \citet{Jebaraj2024} reported that this shock was among the fastest in situ shocks ever recorded (\(V_\mathrm{sh} \sim 3000\)~km s\(^{-1}\)).

{The eruption’s source region remains uncertain due to limited far-side imaging. However, \citet{Dresing2025} identified two candidate active regions (ARs) by analysing full-disk observations before and after the event. One AR (AR1) was located in the northern hemisphere, initially designated AR13229 during Carrington rotation 2267 and reappearing as AR13258 in the subsequent rotation. The other AR (AR2), in the southern hemisphere, was originally AR13236 and later evolved into a group of ARs (AR13256, AR13257, and AR13259) during Carrington rotation 2268. Stonyhurst coordinates place these ARs at approximately \(-109^\circ\) (AR1) and \(-169^\circ\) (AR2) longitude at the time of eruption. The estimated positions of these ARs are marked in Fig.~\ref{fig:sc_constellation}, along with the locations of spacecraft at the time of the event. For a detailed analysis of AR identification and evolution, see \citet{Dresing2025}.}

Figure~\ref{fig:coronagraph} presents coronagraph observations of the CME and the CME-driven shock wave in white light, taken by the Large Angle and Spectrometric Coronagraph Experiment \citep[LASCO;][]{brueckner95} C2 camera on board the SOlar and Heliospheric Observatory \citep[SOHO;][]{domingo95} and the COR2 coronagraph \citep{howard2008} on board the Solar TErrestrial RElations Observatory Ahead \citep[STEREO-A;][]{kaiser08} spacecraft. The event likely involved two simultaneous CMEs on the Sun’s far side --- one moving north and the other south (see panels a and b) --- originating from AR1 and AR2, respectively. Both CMEs were enveloped by a bright white-light front identified as the outer shock wave. This shock is clearly visible in the coronagraph images, and the deflection of several streamers on the west limb (see S1D and S2D in panels d–f) confirms a wide disturbance propagating in the corona.

The shock was also linked to an intense, circumsolar SEP event \citep[see][]{Dresing2025}, with particle intensity increases recorded at Parker Solar Probe, the first Sun–Earth Lagrange point (L1), Solar Orbiter \citep[][]{muller20}, STEREO-A, BepiColombo \citep[][]{benkhoff21}, and possibly even Mars. {Notably, Earth’s nominal Parker spiral connection lies about 120\(^{\circ}\) in longitude from AR2 (Fig.~\ref{fig:sc_constellation}), yet it still experienced a rapid SEP onset. According to \citet{Dresing2025}, the observed long-lasting anisotropy at L1 suggests either very efficient particle transport across large longitudinal separations close to the Sun or a direct magnetic connection to a wide shock wave, likely the eastern flank of the CME erupting from AR2.}
Furthermore, each spacecraft eventually recorded an energetic storm particle (ESP) event when {a shock} arrived, despite the broad range of heliolongitudes involved. As shown in the sections that follow, {a possible explanation is that the western flank of the shock wave originating from AR1 propagated to the far-side spacecraft, suggesting that the IP shock associated with the 13 March 2023 event was not only exceptionally fast but also strikingly wide.}

\section{{EUHFORIA  setup for the 13 March 2023 event}}\label{sec:set-up}
\begin{table*}[t]
\begin{center}
\caption{Parameters of the CMEs and the shock wave injected into the EUHFORIA simulation.}
\label{tab:CME_pars}

\begin{tabular}{cccccccc}
\hline
\multicolumn{8}{c}{{Pre-event ellipsoid cone CMEs}} \\
\hline
Insertion & Lat & Lon & $\omega_{\text{major}}$ & $\omega_{\text{minor}}$ & Tilt Angle &  Speed &  \\
 Date \& Time & [deg] & [deg] & [deg] & [deg] & [deg] & [km~s\(^{-1}\)] & \\
\hline
2023-03-10 19:29 & -3  &  50  & 43  & 18 & -3  & 502 & \\
2023-03-10 23:45 & -31 &  28  & 51  & 19 & -80 & 580 & \\
2023-03-11 22:11 & -50 &  38  & 35  & 15 & 1   & 638 & \\
2023-03-12 09:49 & 35  & -111 & 54  & 20 & 68  & 600 & \\
2023-03-13 00:05 & -42 &  2   & 44  & 22 & -82 & 781 & \\
\hline

\\[0.05cm] 

\multicolumn{8}{c}{{Post-event ellipsoid cone CME}} \\
\hline
Insertion & Lat & Lon & $\omega_{\text{major}}$ & $\omega_{\text{minor}}$ & Tilt Angle &  Speed &  \\
 Date \& Time & [deg] & [deg] & [deg] & [deg] & [deg] & [km~s\(^{-1}\)] & \\
\hline
2023-03-13 13:59  & 33 &  31   & 39  & 19 & 55 & 1179 & \\
\hline
\\[0.05cm] 

\multicolumn{8}{c}{{Main event shock waves}} \\
\hline
Insertion & Lat & Lon & $\omega$ &  Shock speed & && \\
 Date \& Time & [deg] & [deg] & [deg] & [km~s\(^{-1}\)]&&& \\
\hline
2023-03-13 05:00 &  15 & -169 & 110 &{3000} & & &\\
2023-03-13 05:00 & -12 & -109 & 110 &{3000} & & &  \\
\hline

\\[0.05cm]

\multicolumn{8}{c}{{Main event spheromak CMEs}} \\
\hline
 Insertion & Lat & Lon & Radius & Tilt Angle &  Speed & Hel. Sign  & Flux \\
 Date \& Time& [deg] & [deg] & [R\(_s\)] & [deg] & [km~s\(^{-1}\)] & [\(\pm 1\)]  & [Wb] \\
\hline
2023-03-13 05:15 & 15  & -169 & 12 & 90.0 & 1800.0  & 1.0  & $5 \times 10^{13}$ \\
2023-03-13 05:15 & -12 & -109 & 12 & 90.0  & 1800.0  & 1.0  & $5 \times 10^{13}$ \\
\hline
\end{tabular}
\end{center}
\footnotesize{ \textbf{Notes.} The first section lists pre-event ellipsoid cone CMEs, with their insertion date and time at 0.1 au, latitude (Lat), longitude (Lon), major and minor angular half widths ($\omega_{\text{major}}$, $\omega_{\text{minor}}$), tilt angle, and insertion speed. The second section provides the details of a single post-event ellipsoid cone CME. The third section specifies the injection time, location, half width $\omega$ and speed of the shock waves. The final section describes the main-event spheromak CMEs, with the spheromak-specific parameters being its radius, helicity sign, and magnetic flux. Both the spheromak and the ellipsoid CMEs are injected with a uniform temperature of $8$~MK and density $\rho =10^{-18}$~kg~m$^{-3}$.} 
\end{table*}

To investigate the {potential} role of a wide shock wave in the 13 March 2023 eruption, we performed a simulation using the EUHFORIA model \citep[][]{pomoell18,Poedts2020}. EUHFORIA consists of two primary components: a coronal model that derives solar wind and magnetic field conditions at 0.1 au from photospheric magnetograms, and a heliospheric model that propagates the plasma and magnetic structures to larger radial distances by solving the ideal MHD equations.
In this work, we employed the Air Force Data Assimilative Photospheric Flux Transport \citep[ADAPT;][]{arge2010, hickmann2015} model in conjunction with Global Oscillation Network Group \citep[GONG;][]{harvey96} magnetograms to specify the coronal conditions. Specifically, we used the first GONG-ADAPT realization produced at 00:00 UT on 13 March 2023, and then set the MHD boundary conditions at 0.1~au following \citet{pomoell18}, with two modifications:  
(i) Increasing the reference magnetic field at 0.1~au from 300~nT to \(B_{\text{fsw}} = 500\)~nT, to achieve better agreement with in situ spacecraft data.  
(ii) Raising the minimum solar wind speed from 275 to 350~km~s\(^{-1}\), ensuring that the solar wind remains superfast at the inner boundary (0.1~au), as required by the MHD module of EUHFORIA \citep[e.g.][]{Samara2024}. Our computational domain extends to 3~au, with a radial grid spacing of \(\Delta r = 0.6\,R_s\) and an angular resolution of 1\(^{\circ}\) in both latitude and longitude. 

To better capture the heliospheric conditions prior to the eruption of the primary CMEs on 13 March 2023, our EUHFORIA simulation includes five pre-event CMEs and one post-event CME. These CMEs were selected based on their potential impact on the heliospheric environment surrounding the main event. Specifically, we included CMEs that erupted within three days before the main event and met at least one of the following criteria: (i) Earth-directed CMEs with projected speeds exceeding 350 km~s\(^{-1}\), or (ii) CMEs with angular widths greater than 90\(^{\circ}\) and speeds above 500~km~s\(^{-1}\), as listed in the CDAW SOHO LASCO CME catalogue\footnote{\url{https://cdaw.gsfc.nasa.gov/CME_list/}} \citep{Yashiro2004}. The kinematic parameters of these CMEs are detailed in Table~\ref{tab:CME_pars}. To better constrain these parameters, we applied the graduated cylindrical shell model \citep{Thernisien2006GCS,Thernisien2011} to multi-point coronagraph observations from STEREO-A and SOHO, reconstructing the CME properties from approximately 3.5 to 20~\(R_\odot\).  

{To model the 13~March~2023 eruption, we injected two spheromak CMEs at 0.1~au at the inferred locations of AR1 and AR2 (see Fig.~\ref{fig:sc_constellation}). Each spheromak was initialized with a radius of 12~\(R_\odot\), corresponding to a ${\sim}$30\(^{\circ}\) half-width. 
Prior to the spheromak injections, we introduced two wide shock waves at the inner boundary by applying Rankine–Hugoniot (RH) jump conditions, as described in Appendix~\ref{app:EUHFORIA}. 
These shocks were assigned a half-width of 110\(^{\circ}\) and a uniform speed of 3000~km~s\(^{-1}\), consistent with estimates by \citet{Jebaraj2024}. The central region of the wave fronts remains CME-driven, while the flanks propagate freely. The shock centres, like in the spheromaks, coincide with the estimated AR locations and are placed 60\(^{\circ}\) apart, resulting in significant overlap between their wave fronts. In these overlapping regions, the shocks are treated as a single wave, determined by the shock speed and local upstream solar wind conditions. This setup aligns with coronagraph observations, which suggest a shock envelope encompassing both CMEs (Fig.~\ref{fig:coronagraph}).  The total azimuthal extent of the injected shock waves is 280\(^{\circ}\), as shown in Fig.~\ref{fig:sc_constellation}.}

{We adopted a 110\(^{\circ}\) half-width for the shock from AR1, ensuring that the western tip of the shock flank travels in the direction of L1. Applying the same half-width to the shock from AR2 places its eastern flank at 81\(^{\circ}\) in azimuthal angle (Stonyhurst coordinates), which remains approximately 10\(^{\circ}\) away from L1's nominal magnetic connection to the solar surface \citep[see][or Fig.~\ref{fig:sc_constellation}]{Dresing2025}.
Given the observed anisotropic SEP event onset at L1, this suggests that the shock from AR2 may have been even wider than assumed. 
Moreover, while we impose a constant shock speed of 3000~km~s$^{-1}$ across the entire wave front for simplicity, the actual speed is likely lower at the flanks than at the CME-driven centre. Nevertheless, this setup allows us to investigate the evolution of wide shock flanks in a preconditioned heliosphere, gain further insights into the shock’s large-scale structure, and make comparisons with observations from multiple spacecraft.}

\begin{figure*}
    \centering
    \includegraphics[width=0.49\linewidth]{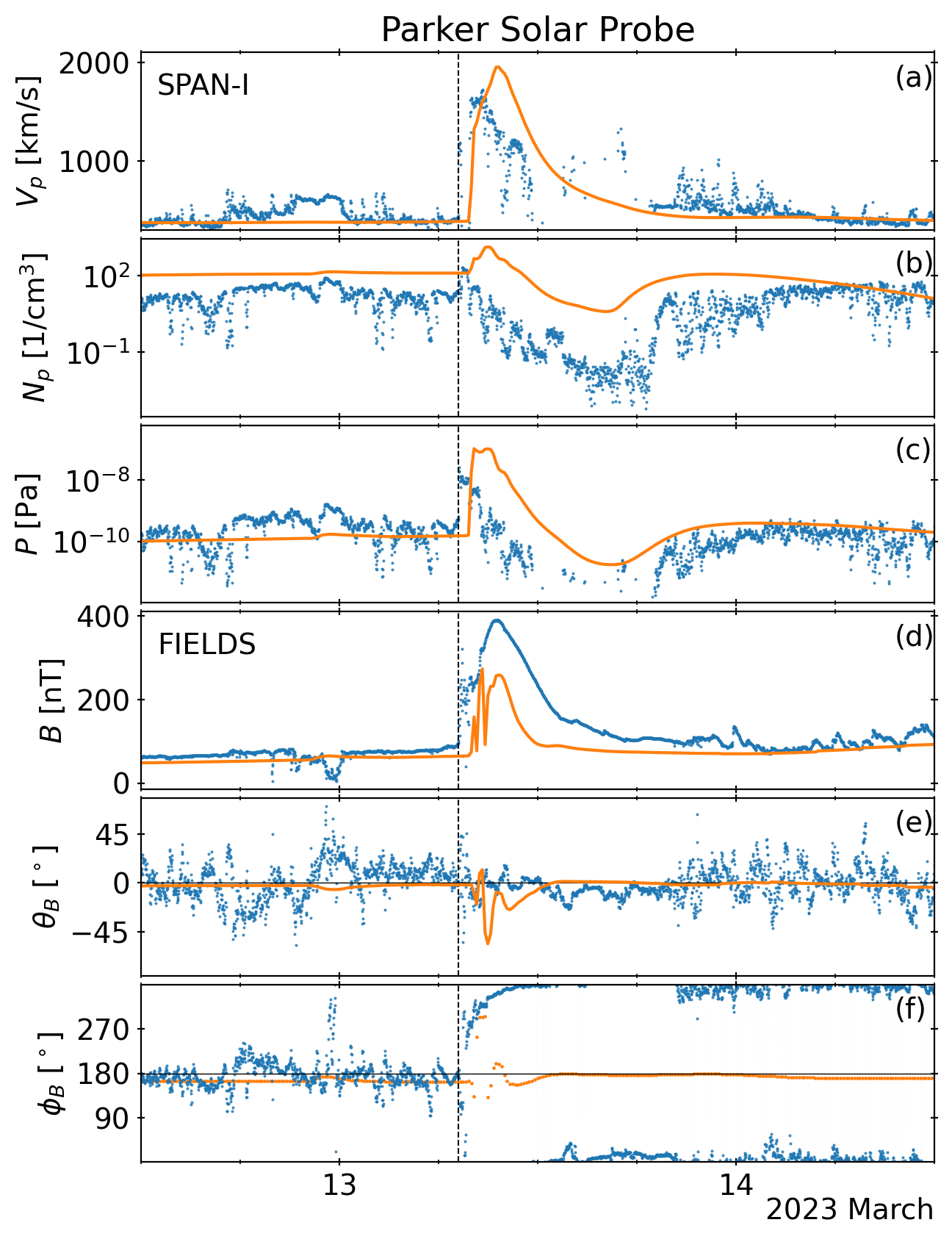}
    \includegraphics[width=0.49\linewidth]{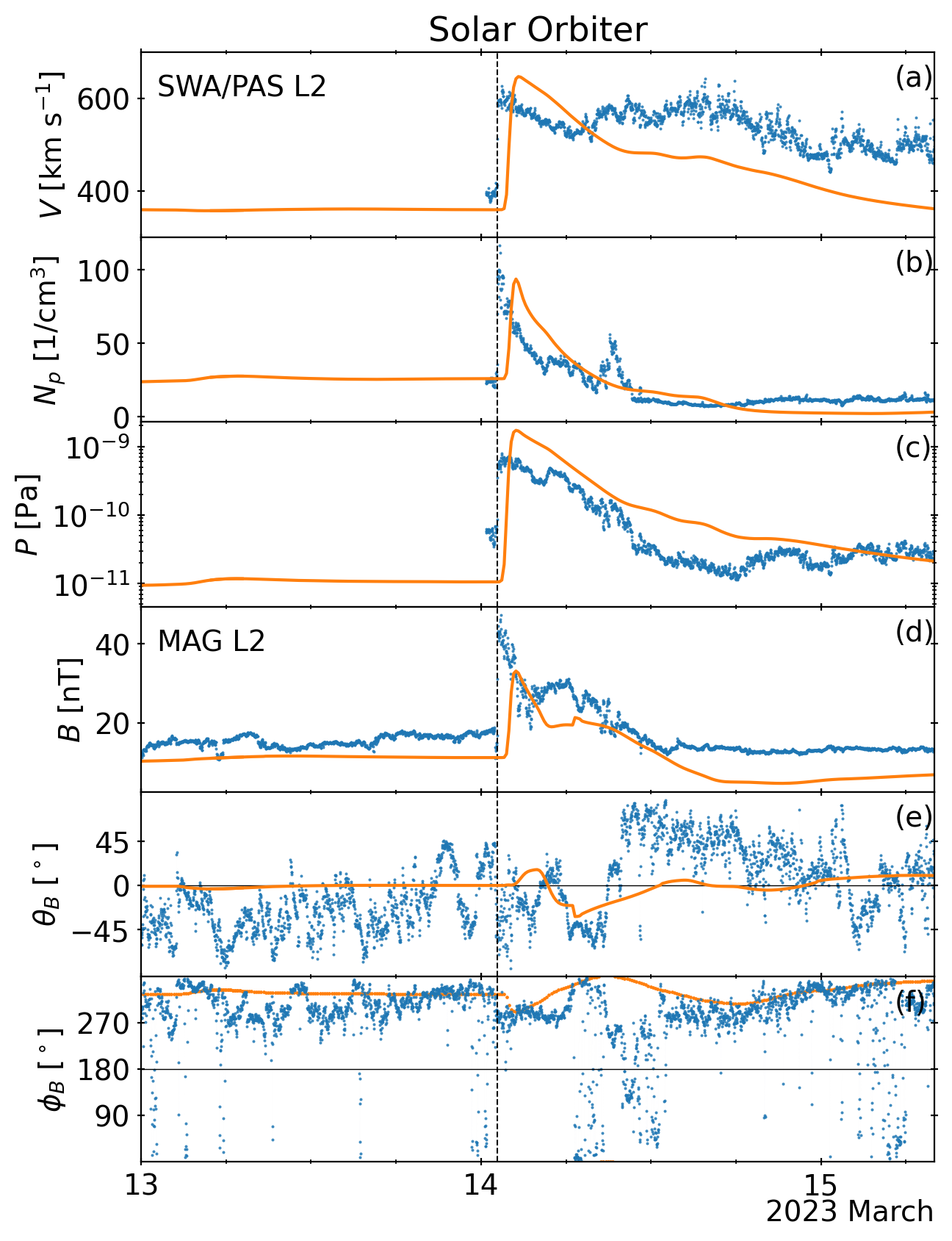}\\
    \includegraphics[width=0.49\linewidth]{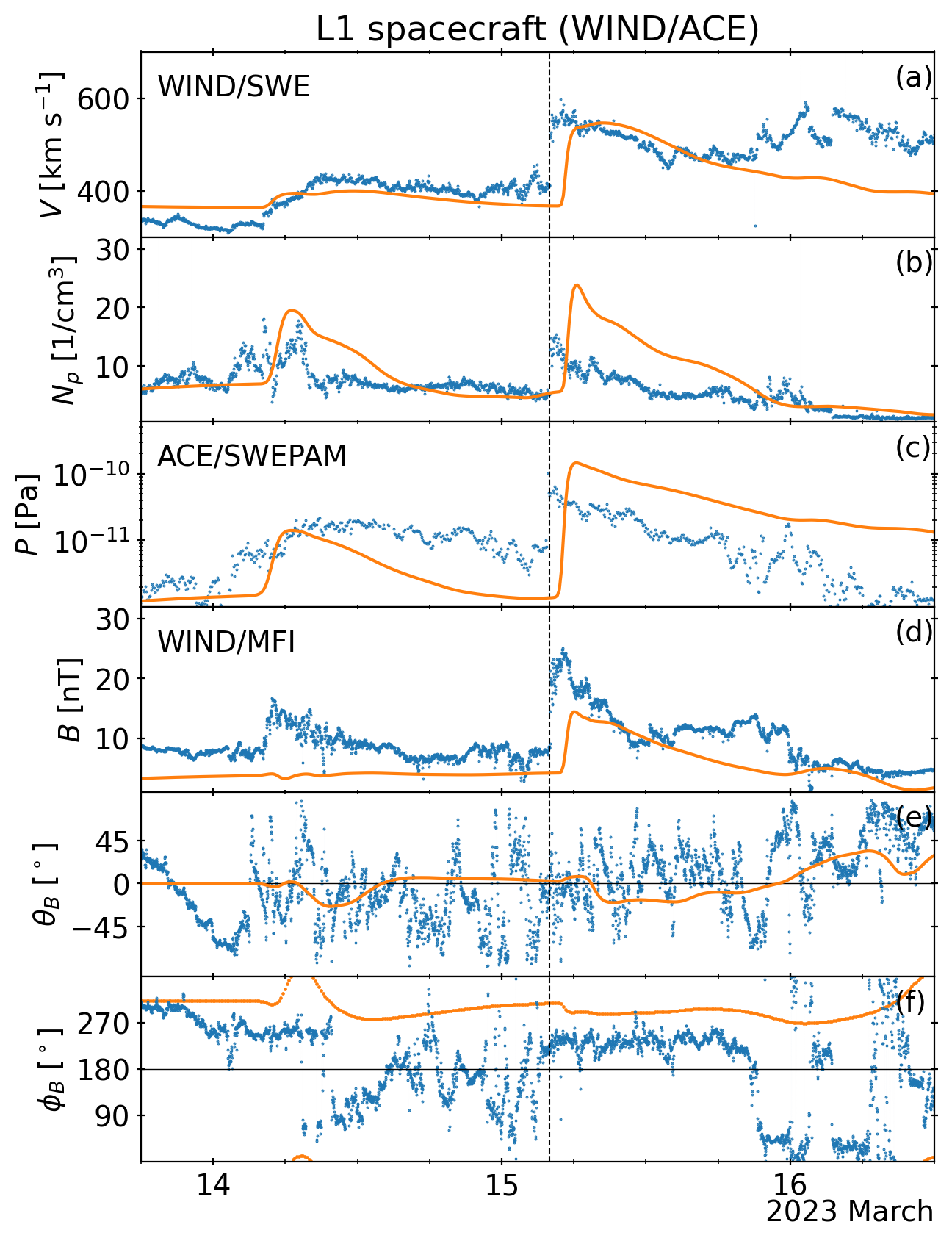}
    \includegraphics[width=0.49\linewidth]{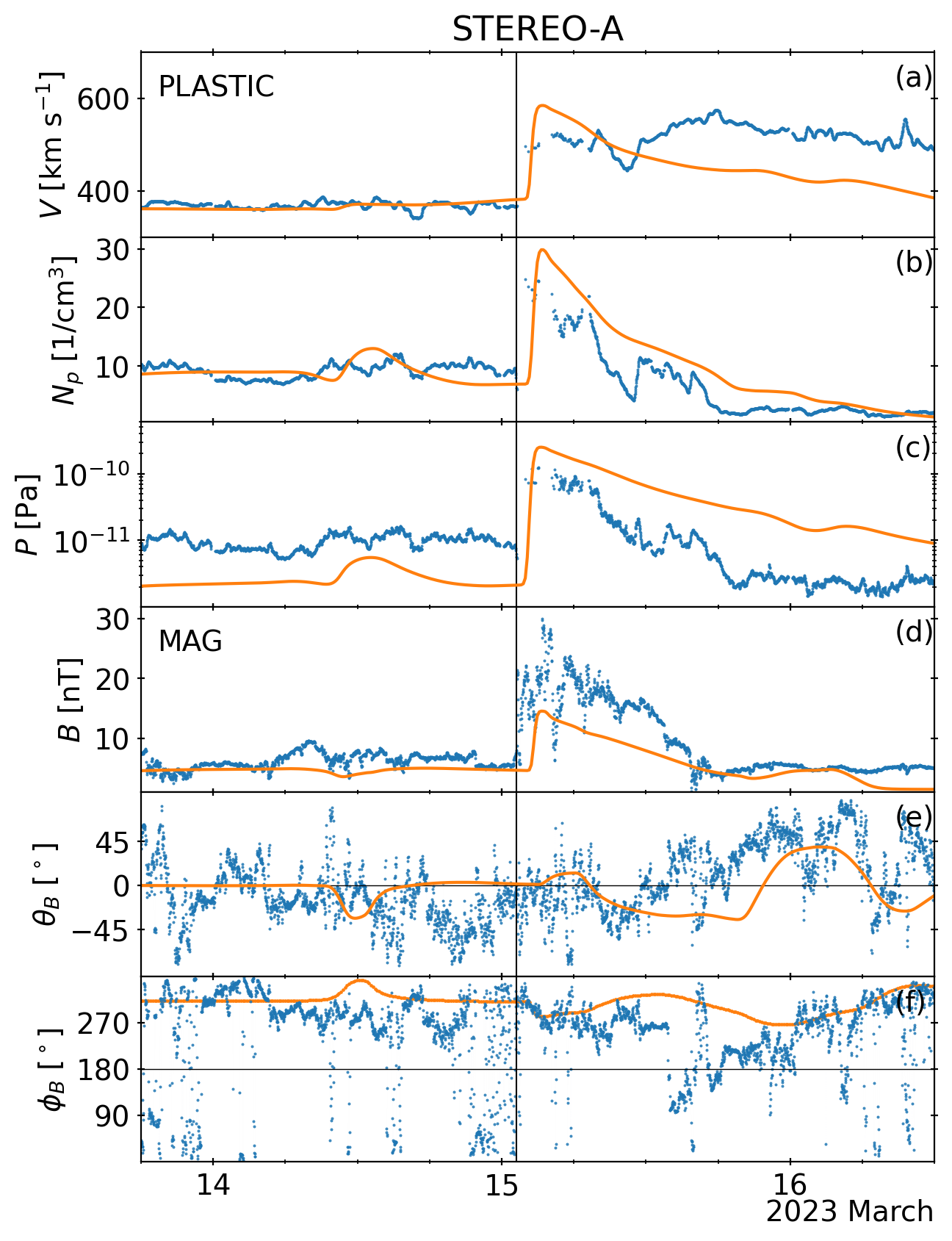}   
    \caption{Comparison of observed (blue) and simulated (orange) solar wind parameters at various spacecraft locations: Parker Solar Probe (top left), Solar Orbiter (top right),  L1 (ACE and WIND; bottom left), and STEREO-A (bottom right). For each spacecraft, the panels display: (a) solar wind speed, (b) proton number density, (c) thermal proton pressure, (d) magnetic field magnitude, (e) magnetic latitudinal angle, and (f) magnetic azimuthal angle. }\label{fig:euh obs}
\end{figure*}

\begin{figure*}
\centering
    \includegraphics[width=0.85\linewidth]{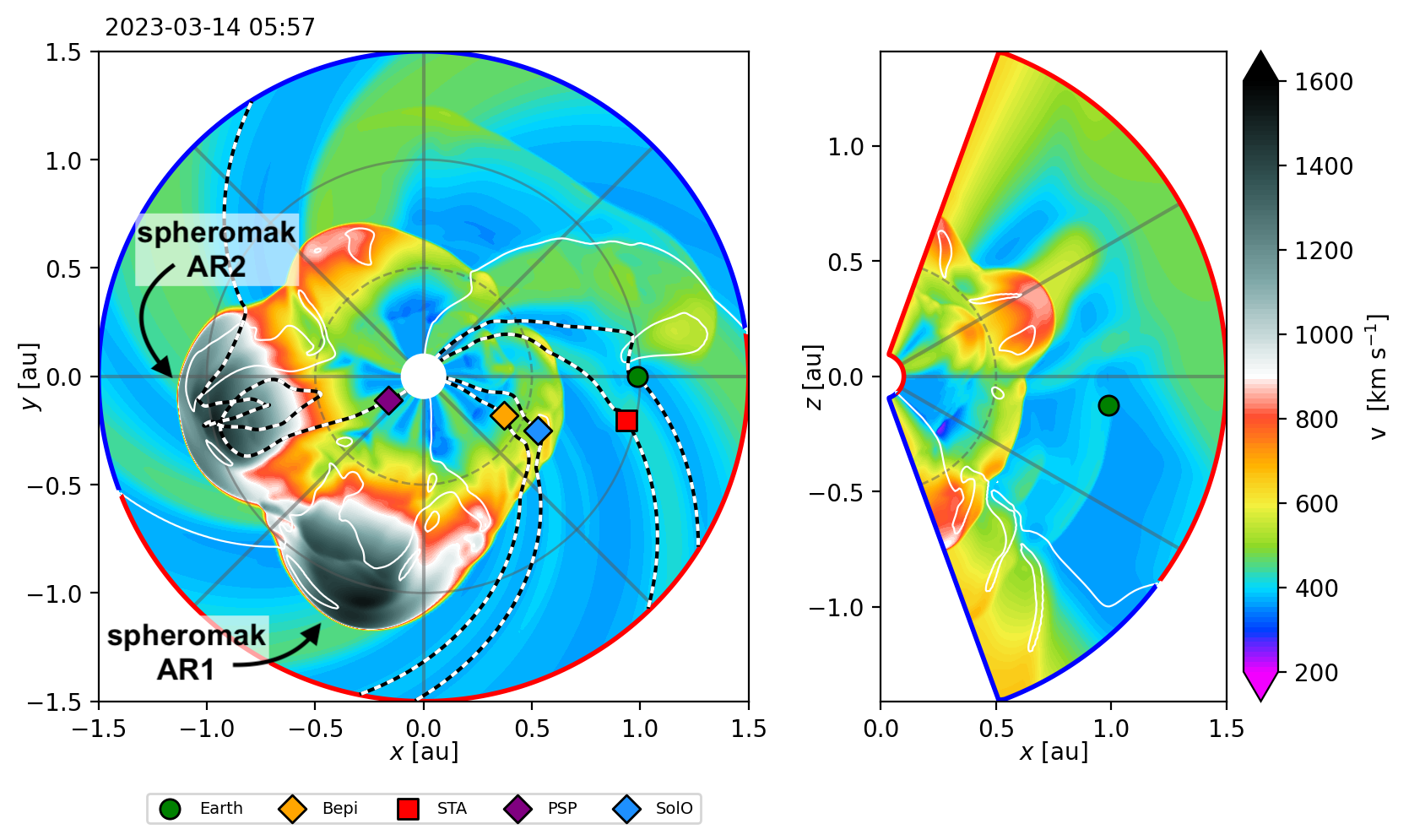}\\
    \includegraphics[width=0.85\linewidth]{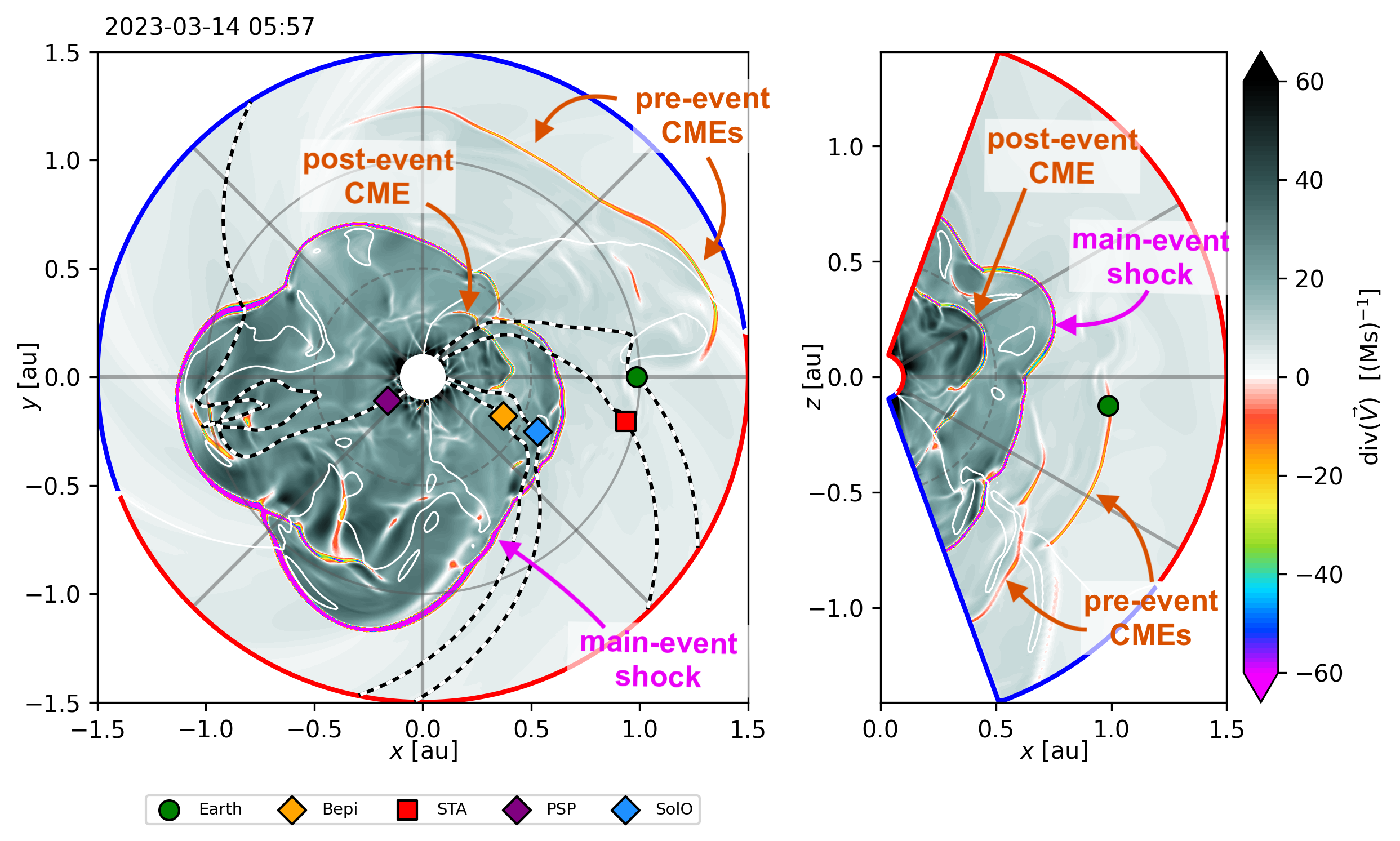}
    \caption{Snapshot of the EUHFORIA simulation on 14 March at 05:57 UT. Left panels: Solar equatorial plane. Right panels: Meridional slices containing Earth. Top row: Solar wind speed. Bottom row: Divergence of the solar wind velocity $\nabla\cdot\vec{V}_{\text{sw}}$. Dashed lines indicate projections of magnetic field lines connecting to various spacecraft. A video version of this figure is available. }\label{fig:euh slices}
\end{figure*}

\section{Simulation results}\label{sec:results}

Figure~\ref{fig:euh obs} presents a comparison between the EUHFORIA simulation results and observations from various spacecraft, including Parker Solar Probe, Solar Orbiter, STEREO-A, and spacecraft at L1. This figure displays a range of in situ measurements alongside the corresponding quantities from the EUHFORIA simulation. From top to bottom, each panel shows the solar wind proton speed, the proton number density, the thermal pressure, {the magnetic field magnitude,  the Geocentric Solar Ecliptic (GSE) latitudinal magnetic field angle, and the GSE azimuthal magnetic field angle.}
The arrival of the shock wave is clearly identifiable at each spacecraft by a sudden jump in most of these panels and is marked by a dotted line. Comparing these in situ observations with the EUHFORIA simulation results, we find a remarkable agreement. Both the arrival time and the amplitude of the  shock wave align well across all spacecraft.

{An important observation from the multi-spacecraft analysis of in situ data is that, apart from Parker Solar Probe, no other spacecraft detected clear signatures of a flux rope downstream of the shock. This not only aligns well with our simulation results but also supports the hypothesis of a shock flank encounter without any CME signatures to follow.}
{A typical characteristic of a flux rope is a smooth and coherent rotation in the magnetic field angles, yet no such structure is evident in the in situ magnetic field data. Solar Orbiter is the only spacecraft to exhibit some rotation in the magnetic field at the end of the enhanced field region; however, this coincides with an increase in plasma density, which is not a typical flux-rope signature. Furthermore, the enhanced magnetic field strength observed downstream of the shock at multiple spacecraft, along with fluctuations in the magnetic field direction, is expected in a turbulent shock sheath and does not necessarily indicate the presence of a flux rope.}

This is further supported by the EUHFORIA simulation, which, for example, shows variations in the magnetic field angles at STEREO-A, despite no magnetic cloud passing the spacecraft in the simulation. {Moreover, the fact that the simulated magnetic field agrees relatively well with observations — both in magnitude and duration — at spacecraft encountering the modelled shock flank further suggests that no flux rope crossed these spacecraft. This strengthens the interpretation that the observed magnetic field variations are a consequence of the turbulent shock sheath rather than the passage of an underlying magnetic ejecta structure. Alternatively, these variations could result from glancing encounters with the less-organized flanks of the ejecta, where the magnetic field structure is more complex and lacks the coherent rotation expected for a magnetic cloud \citep[e.g.][]{2022Rodriguez-Garcia}.}

Figure~\ref{fig:euh slices} shows a snapshot from the EUHFORIA simulation on 14 March at 05:57 UT, {illustrating the solar wind speed (top row)} and the divergence of solar wind velocity, \(\divV\) (bottom row). The left column displays the solar equatorial plane, while the right column shows a meridional slice passing through Earth. In the solar equatorial plane, the  shock wave is distinctly visible as a pronounced jump in speed, encircling the plot’s centre. Two spheromaks representing the main event appear as extended regions of enhanced velocity, concentrated primarily in the third quadrant. In this study, we do not focus on these spheromak CMEs but instead analyse the freely propagating wave, also visible in the other three quadrants of the EUHFORIA simulation.

The bottom row of Fig.~\ref{fig:euh slices} displays \(\divV\) and highlights the main-event shock wave. Here, \(\divV < 0\) represents converging flows, typical of compression regions or shock fronts, while \(\divV > 0\) indicates diverging flows, characteristic of the expanding solar wind. Downstream of the main-event shock, the plasma is characterized by  \(\divV \gg 0\), indicating a strongly expanding flow.  

{At the time of the snapshot shown in Fig.~\ref{fig:euh slices}, the main-event shock had already overtaken two earlier CMEs, launched on 12 and 13 March, respectively (see Table~\ref{tab:CME_pars}). Initially, the shock spanned 280\(^{\circ}\) in longitude, from 1\(^{\circ}\) to -281\(^{\circ}\) (\(=81^{\circ}\)). However, due to the lateral expansion of its flanks, the disturbance has since become quasi-circumsolar in the equatorial plane, with both the eastern and western flanks expanding by over ${\sim}$40\(^{\circ}\) and thus merging around 40\(^{\circ}\) longitude.   Additionally, our simulation likely underestimates the lateral expansion of the shock, as it is injected with a purely radial velocity at 0.1~au. As a result, a narrower initial shock half-width, such as 70\(^{\circ}\) instead of 110\(^{\circ}\) from AR1, should, in principle, still have allowed the shock to reach L1.} 

In addition to the main-event shock, several other shock waves are visible, associated with preceding CMEs included in the simulation (see Table~\ref{tab:CME_pars}). For example, the meridional slice in Fig.~\ref{fig:euh slices} shows a pre-event CME whose northern flank is expected to pass Earth around 06:00 UT on 14 March (see also Fig.~\ref{fig:euh obs}). In the solar equatorial plane, the three earliest pre-event CMEs have merged into a single, broad compression front. This type of merging is reminiscent of what likely occurred during the main event itself, when two fast CMEs erupted almost simultaneously, producing the exceptionally wide shock wave analysed in this work. 

Finally, the post-event CME, propagating through the shocked solar wind behind the main-event shock, is also visible in Fig.~\ref{fig:euh slices}. As shown more clearly in the video version of the figure, this post-event shock initially catches up with the main-event CME, but ultimately dissipates before ever overtaking the main-event shock.

Focusing on the circumsolar wave, three key characteristics stand out from this 3D analysis:  
(i) the freely propagating flanks travel significantly slower than the central, spheromak-driven portion of the shock,  
(ii) the wave front exhibits considerable non-uniformity and deformation along its extent, and  
(iii) it is predominantly characterized by \(\divV \ll 0\), except at the very tips of the flanks, suggesting that it likely persists as a shock wave rather than transitioning into a simple compression wave.

\begin{figure}
\centering
    \includegraphics[width=0.95\linewidth]{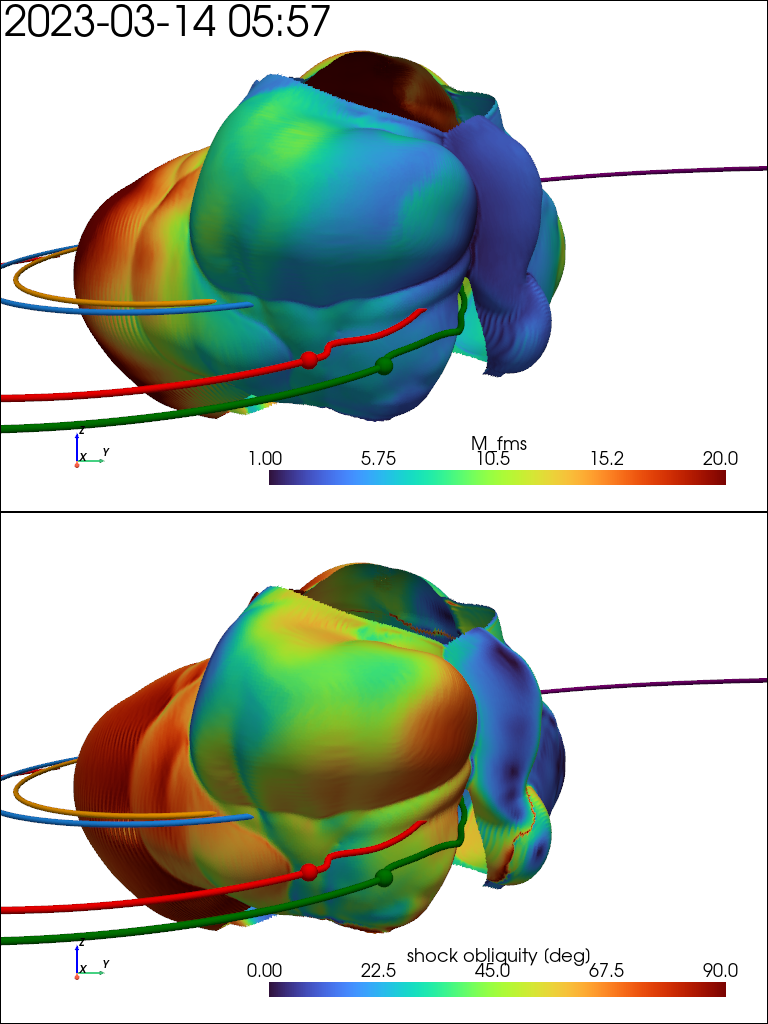}
    \caption{Main-event shock surface, along with the magnetic field lines connecting Earth (green), STEREO-A (red), Solar Orbiter (blue), BepiColombo (orange), and Parker Solar Probe (purple). 
    In the top panel, the shock surface is colour-coded by the fast magnetosonic Mach number (\( M_{\text{fms}} \)), while in the bottom panel, it is colour-coded by the shock obliquity angle (\( \theta_{B_n} \)). 
    The associated video provides a 360\(^{\circ}\) view around the shock surface.}\label{fig:shock}
\end{figure}

To confirm that the disturbance is a shock wave, we employ the shock wave tracer described in \citet{wijsen22} and calculate its properties such as the fast-magnetosonic Mach number, \( M_{\text{fms}} \). The top panel of Fig.~\ref{fig:shock} shows the shock surface on 14 March at 05:58 UT, colour coded according to \( M_{\text{fms}} \). As indicated, \( M_{\text{fms}} > 1 \) throughout, verifying that the wave is indeed a shock. A 360\(^{\circ}\) video of the shock is also available online. As expected, the video reveals that the shock is strongest -- i.e. characterized by the highest \( M_{\text{fms}} \) -- where the spheromaks are driving the wave. However, even at the flanks, where the wave is not driven by the spheromaks, the wave remains a shock until the end of the simulation, at which point the shock has travelled beyond 2~au.

The pronounced deformation of the shock results  from the non-uniform upstream solar wind, a phenomenon previously discussed for CME-driven shocks in, for example, \citet{Wijsen2023}. This deformation gives rise to a shock with highly variable geometry, as demonstrated in the bottom panel of Fig.~\ref{fig:shock}. This panel shows the shock obliquity $\theta_{B_n}$, defined as the angle between the shock normal and the upstream magnetic field. Shock obliquity is believed to play a crucial role in the acceleration of energetic particles. For example, diffusive shock acceleration is thought to be most efficient for quasi-parallel shocks, where the obliquity angle is near 0\(^{\circ}\) \citep[e.g.][and references therein]{malkov01}.

\begin{figure}
\centering
    \includegraphics[width=0.9\linewidth]{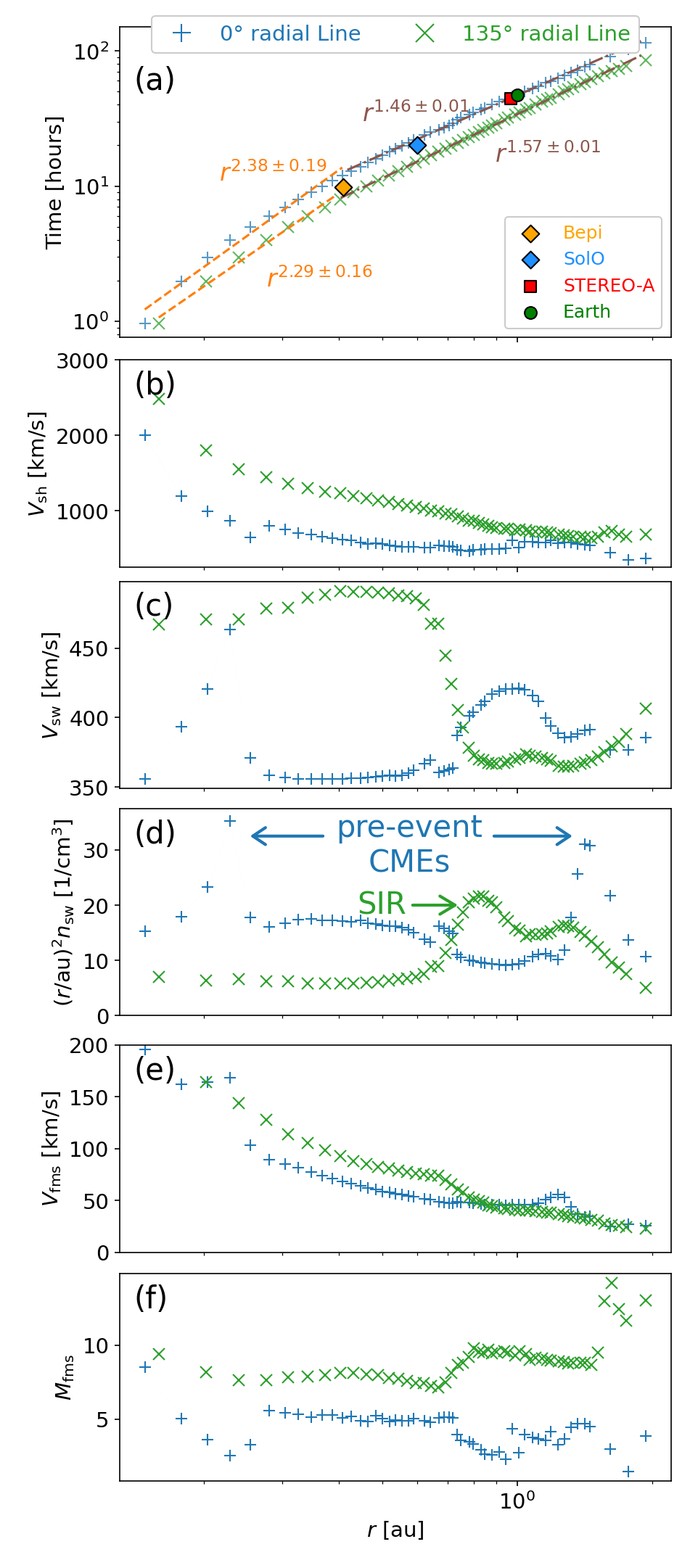}
    \caption{ Evolution of the main-event shock flanks along two radial directions in the solar equatorial plane. The blue symbols represent the line at $0^\circ$, aligned with the positive x-axis in Fig.~\ref{fig:euh slices}, while the green symbols represent a line rotated $135^\circ$ anti-clockwise from the positive x-axis. The  panels present: (a) hours elapsed since the blast wave injection, including power-law fits and observed spacecraft shock crossing times; (b) shock speed; (c) upstream solar wind speed; (d) upstream solar wind density, scaled with $r^2$; (e) upstream fast-magnetosonic speed;  and (f) fast-magnetosonic Mach number ($M_{\text{fms}}$). }\label{fig:euh cob}
\end{figure}

To further illustrate the evolution of the freely propagating shock wave, Fig.~\ref{fig:euh cob} shows its progression along two radial directions in the solar equatorial plane. The first direction (blue symbols), at a longitude of $0^\circ$ degrees, aligns with the x-axis in Fig.~\ref{fig:euh slices}. The second direction (green symbols), at $135^\circ$ longitude, corresponds to the radial line in the second (upper left) quadrant of the solar equatorial plane in Fig.~\ref{fig:euh slices}. The $135^\circ$ direction was selected because no pre-event CME propagates along this path, meaning that the shock interacts only with the unperturbed solar wind.

Figure~\ref{fig:euh cob}a presents the time elapsed since the initial blast wave injection on 13 March at 05:00~UT as a function of radial distance from the Sun, with observed shock crossing times for various spacecraft also indicated. These observations align well with the simulation, reinforcing the accuracy of the model. The panel also includes power-law fits of the form \( t \propto r^{\alpha} \). Near the model’s inner boundary (0.1–0.4 au), we find power-law exponents of  
\( r^{2.358 \pm 0.19} \)  
and  
\( r^{2.29 \pm 0.17} \)  
for the $0^\circ$ and $135^\circ$ trajectories, respectively. Beyond 0.4 au, the expansion slows, with exponents of  
\( r^{1.46 \pm 0.01} \)  
and  
\( r^{1.57 \pm 0.01} \)  
for the same respective trajectories.  These values are consistent with the analytical behaviour of freely propagating pressure waves. In an idealized scenario where energy is deposited instantaneously (e.g. an explosion), the shock initially decays rapidly before transitioning into a regular pressure wave. {This problem, well-studied in uniform hydrodynamic media, is described by the Sedov–von Neumann–Taylor (SVT) solution \citep[][]{Sedov1946,taylor1950I,taylor1950II,vonNeumann1963}, which predicts \( t \propto r^{3/2} \) for a medium where density decreases as \( 1/r^2 \) (see also \citealt{Cavaliere1976} and our Appendix~\ref{app:sedov}).
Similar dependences for freely propagating shock stages in the solar wind have been identified in previous studies. \citet{Smart1985} examined the evolution of IP shocks and found that once a CME-driven shock ceases to be actively driven by the ejecta, its propagation follows a power-law decay akin to the SVT blast wave solution. \citet{Pinter1990} further analysed decoupled IP shocks and demonstrated that their radial evolution can be described by self-similar solutions, with a transition from a piston-driven phase to a freely propagating state.
However, the SVT solution assumes a uniform, expanding upstream plasma with negligible pressure compared to the downstream, so an exact match is not expected. Notably, near the inner boundary, particularly along the $0^\circ$ trajectory, the slope is steeper due to interaction with a pre-event CME.}

Panels (c) and (d) of Fig.~\ref{fig:euh cob} show the variability in the upstream solar wind speed and density, respectively. Along the $135^\circ$ line, a significant density enhancement occurs around 0.75~au, corresponding to stream interaction regions encountered by the shock wave as it propagates outwards. Along the $0^\circ$ radial line, two prominent density peaks are evident near 0.2~au and 1.25~au, associated with two pre-event CMEs overtaken by the freely propagating shock.

{The evolution of the fast magnetosonic speed, \( V_{\text{fms}} \), and Mach number, \( M_{\text{fms}} \), is shown in panels (e) and (f), respectively. Notably, \( V_{\text{fms}} \) decreases almost monotonically with increasing heliocentric distance,  a trend that plays a crucial role in sustaining the shock wave as it propagates outwards. Since solar wind plasma that was initially shocked at smaller radial distances encounters progressively lower magnetosonic speeds, it can continue to reinforce the shock structure at larger distances without requiring additional energy input. As a result, shocks that appear relatively weak near the Sun can seem stronger farther out purely due to the evolving solar wind environment \citep[see also][]{Echer2019}. }

{This effect is particularly pronounced along the 135\(^{\circ}\) trajectory, where \( M_{\text{fms}} \) increases significantly around 0.8 au and 1.6~au. The first increase is driven by the sharp decrease in \( V_{\text{fms}} \) as the shock encounters a stream interaction region. The second increase is resulting from the shock propagating into a low-density rarefaction region trailing a fast wind stream.  Finally, as shown in panel (f), \( M_{\text{fms}} \) consistently exceeds unity along both radial lines, confirming that the disturbance remains a shock wave throughout its propagation from the upper corona to 2~au.}

\section{Conclusions}\label{sec:conclusions}

{In this study we investigated how the flanks of a wide CME-driven shock can persist} in the heliosphere as freely propagating waves, using the EUHFORIA MHD solar wind model. 
We modified EUHFORIA to allow the direct injection of shock waves with specified speeds at the inner boundary by ensuring that their parameters meet the RH jump conditions for the given shock speed.
Our key findings are summarized as follows:

\begin{enumerate}

\item {Sustained, freely propagating shock flanks:} 
EUHFORIA simulations show that the flanks of wide CME-driven shocks can persist beyond 2\,au, even in the absence of a continuous CME driving at those flanks. {A key factor in this persistence is the decreasing fast magnetosonic speed with heliocentric distance, allowing initially shocked solar wind to sustain the shock farther out without additional energy input.}
\item {Application to the 13~March~2023 event:} 
We modelled two nearly simultaneous CMEs that drive a single, wide shock. In this configuration, the central shock region near the apex remains CME-driven, whereas the flanks propagate freely. The resulting global shock front, initially injected with a 280$^\circ$ angular width, extends into a quasi-circumsolar wave. The modelled shock arrival times and amplitudes of key plasma parameters (e.g. speed and density) closely match in situ observations from multiple spacecraft spanning various radial distances and longitudes. 

{The 13 March 2023 event represents an extreme case, with shock speeds reaching up to 3000~km~s$^{-1}$ at Parker Solar Probe, leading \citet{Jebaraj2024} to term it an `almost astrophysical' shock. This exceptionally high speed may explain why the shock persisted at such large heliocentric distances. For weaker shocks, the lack of a sustained CME driving at the flanks may lead to a faster dissipation.}

\item {Implications for widespread SEP events:}
The persistence of shock flanks may contribute to the spatial extent and duration  of widespread SEP events. For extreme cases, like the event discussed in this work, the shock likely needs to propagate well into IP space to help explain ESP events observed across a wide longitudinal range \citet{Dresing2025}. Variations in observed particle profiles may thus be influenced not only by transport processes but also by differences in shock geometry and local plasma conditions.

\end{enumerate}

{While our simulation results suggest that the 13 March 2023 event can be explained by a single, wide shock, this is only one possible interpretation of the observations. Alternatively, the event could also result from a shock-flank encounter with a preceding CME \citep[see also][]{Dresing2025}. In the presented EUHFORIA simulation, these earlier CMEs either miss the spacecraft or arrive too late (more than 12 hours after the observed shock).}  

{However, heliospheric MHD models such as EUHFORIA and ENLIL \citep{odstrcil03} struggle to accurately reproduce shock-flank encounters \citep[e.g.][]{Bain2016ApJ,scolini19}. This is partly because they start their simulations at 0.1~au, underestimating the true shock extent by injecting only the CME ejecta while neglecting the shock wave and sheath that form at lower altitudes. As a result, these models also face difficulties in predicting SEP onset times when an observer connects to the shock flank \citep[see e.g.][]{wijsen22}.}  

{By explicitly injecting a shock ahead of the CME ejecta that satisfies the RH jump conditions at the inner boundary, our simulations take a step towards addressing this limitation, providing a more realistic representation of the shock evolution in the heliosphere. In principle, such a shock and sheath should be injected ahead of every sufficiently fast CME. A more comprehensive, though computationally expensive, approach would involve modelling the entire solar eruption from the low corona onwards to fully capture the formation and expansion of the shock \citep[e.g.][]{Manchester2004JGRA,linan2025}.}

{Finally, we note that the results obtained in this study rely on an ideal MHD framework, which does not account for certain physical processes that may influence the persistence of freely propagating shock fronts, such as kinetic-scale dissipation. Future work will focus on investigating effects beyond ideal MHD to further improve our understanding of shock evolution and its role in particle acceleration and transport. } 

\begin{acknowledgements}
N.W.\ acknowledges funding from the Research Foundation -- Flanders (FWO -- Vlaanderen, fellowship no.\ 1184319N) and from the KU Leuven project 3E241013. Computational resources used for the EUHFORIA simulations presented in this work were provided by the VSC (Flemish Supercomputer Center), funded by FWO -- Vlaanderen and the Flemish Government – department EWI.

Work in the University of Turku was performed under the umbrella of Finnish Centre of Excellence in Research of Sustainable Space (FORESAIL) funded by the Research Council of Finland (grant No.\ 352847). I.C.J.\ and N.D.\ are grateful for support by the Research Council of Finland (SHOCKSEE, grant No.\ 346902).
We thank the members of the data analysis working group at the Space Research Laboratory of the University of Turku, Finland for useful discussions.
We acknowledge funding by the European Union’s Horizon 2020 / Horizon Europe research and innovation program under grant agreement No.\ 101004159 (SERPENTINE) and No.\ 101134999 (SOLER). The paper reflects only the authors' view and the European Commission is not responsible for any use that may be made of the information it contains.
I.C.J.\ also acknowledges support from the International Space Science Institute (ISSI) in Bern through ISSI International Team project No.~23-575, ``\textit{Collisionless Shock as a Self-Regulatory System}''. 

A.K. acknowledges financial support from NASA’s NNN06AA01C (80MSFC19F0002 SO-SIS Phase-E and Parker Solar Probe EPI-Lo) contract.

E.P.\ acknowledges support from NASA's PSP-GI (grant No.\ 80NSSC22K0349), HGI (grant No.\ 80NSSC23K0447), LWS (grant No.\ 80NSSC19K0067), and LWS-SC (grant No.\ 80NSSC22K0893) programmes, as well as NSF's PREEVENTS (grant No.\ ICER-1854790) programme.

\end{acknowledgements}

\bibliographystyle{aa} 
\bibliography{references, references_Koulou}

\begin{thebibliography}{69}
\expandafter\ifx\csname natexlab\endcsname\relax\def\natexlab#1{#1}\fi

\bibitem[{{Akhiezer}(1975)}]{Akhiezer75}
{Akhiezer}, A.~I. 1975, {Plasma electrodynamics - Vol.1: Linear theory; Vol.2:
  Non-linear theory and fluctuations}

\bibitem[{{Arge} {et~al.}(2010){Arge}, {Henney}, {Koller}, {Compeau}, {Young},
  {MacKenzie}, {Fay}, \& {Harvey}}]{arge2010}
{Arge}, C.~N., {Henney}, C.~J., {Koller}, J., {et~al.} 2010, in American
  Institute of Physics Conference Series, Vol. 1216, Twelfth International
  Solar Wind Conference, ed. M.~{Maksimovic}, K.~{Issautier},
  N.~{Meyer-Vernet}, M.~{Moncuquet}, \& F.~{Pantellini} (AIP), 343--346

\bibitem[{{Bain} {et~al.}(2016){Bain}, {Mays}, {Luhmann}, {Li}, {Jian}, \&
  {Odstrcil}}]{Bain2016ApJ}
{Bain}, H.~M., {Mays}, M.~L., {Luhmann}, J.~G., {et~al.} 2016, \apj, 825, 1

\bibitem[{{Benkhoff} {et~al.}(2021){Benkhoff}, {Murakami}, {Baumjohann},
  {Besse}, {Bunce}, {Casale}, {Cremosese}, {Glassmeier}, {Hayakawa}, {Heyner},
  {Hiesinger}, {Huovelin}, {Hussmann}, {Iafolla}, {Iess}, {Kasaba},
  {Kobayashi}, {Milillo}, {Mitrofanov}, {Montagnon}, {Novara}, {Orsini},
  {Quemerais}, {Reininghaus}, {Saito}, {Santoli}, {Stramaccioni}, {Sutherland},
  {Thomas}, {Yoshikawa}, \& {Zender}}]{benkhoff21}
{Benkhoff}, J., {Murakami}, G., {Baumjohann}, W., {et~al.} 2021, \ssr, 217, 90

\bibitem[{{Brueckner} {et~al.}(1995){Brueckner}, {Howard}, {Koomen},
  {Korendyke}, {Michels}, {Moses}, {Socker}, {Dere}, {Lamy}, {Llebaria},
  {Bout}, {Schwenn}, {Simnett}, {Bedford}, \& {Eyles}}]{brueckner95}
{Brueckner}, G.~E., {Howard}, R.~A., {Koomen}, M.~J., {et~al.} 1995, \solphys,
  162, 357

\bibitem[{{Cavaliere} \& {Messina}(1976)}]{Cavaliere1976}
{Cavaliere}, A. \& {Messina}, A. 1976, \apj, 209, 424

\bibitem[{{Chevalier}(1982)}]{Chevalier82}
{Chevalier}, R.~A. 1982, \apj, 258, 790

\bibitem[{{Corona-Romero} \& {Gonzalez-Esparza}(2011)}]{Corona-Romero2011}
{Corona-Romero}, P. \& {Gonzalez-Esparza}, J.~A. 2011, Journal of Geophysical
  Research (Space Physics), 116, A05104

\bibitem[{{Corona-Romero} \& {Gonzalez-Esparza}(2012)}]{Corona-Romero2012}
{Corona-Romero}, P. \& {Gonzalez-Esparza}, J.~A. 2012, in IAU Symposium, Vol.
  286, Comparative Magnetic Minima: Characterizing Quiet Times in the Sun and
  Stars, ed. C.~H. {Mandrini} \& D.~F. {Webb}, 159--163

\bibitem[{{Domingo} {et~al.}(1995){Domingo}, {Fleck}, \& {Poland}}]{domingo95}
{Domingo}, V., {Fleck}, B., \& {Poland}, A.~I. 1995, \solphys, 162, 1

\bibitem[{{Dresing} {et~al.}(2012){Dresing}, {G{\'o}mez-Herrero}, {Klassen},
  {Heber}, {Kartavykh}, \& {Dr{\"o}ge}}]{dresing12}
{Dresing}, N., {G{\'o}mez-Herrero}, R., {Klassen}, A., {et~al.} 2012, \solphys,
  281, 281

\bibitem[{{Dresing} {et~al.}(2025){Dresing}, {Jebaraj}, {Wijsen}, {Palmerio,
  E.}, {Rodríguez-García, L.}, {Palmroos, C.}, {Gieseler, J.}, {Jarry, M.},
  {Asvestari, E.}, {Mitchell, J. G.}, {Cohen, C. M. S.}, {Lee, C. O.}, {Wei,
  W.}, {Ramstad, R.}, {Riihonen, E.}, {Oleynik, P.}, {Kouloumvakos, A.},
  {Warmuth, A.}, {Sánchez-Cano, B.}, {Ehresmann, B.}, {Dunn, P.}, {Dudnik,
  O.}, \& {Mac Cormack, C.}}]{Dresing2025}
{Dresing}, N., {Jebaraj}, I.~C., {Wijsen}, N., {et~al.} 2025, \aap, 695, A127

\bibitem[{{Dresing} {et~al.}(2023){Dresing}, {Rodr{\'\i}guez-Garc{\'\i}a},
  {Jebaraj}, {Warmuth}, {Wallace}, {Balmaceda}, {Podladchikova}, {Strauss},
  {Kouloumvakos}, {Palmroos}, {Krupar}, {Gieseler}, {Xu}, {Mitchell}, {Cohen},
  {de Nolfo}, {Palmerio}, {Carcaboso}, {Kilpua}, {Trotta}, {Auster},
  {Asvestari}, {da Silva}, {Dr{\"o}ge}, {Getachew}, {G{\'o}mez-Herrero},
  {Grande}, {Heyner}, {Holmstr{\"o}m}, {Huovelin}, {Kartavykh}, {Laurenza},
  {Lee}, {Mason}, {Maksimovic}, {Mieth}, {Murakami}, {Oleynik}, {Pinto},
  {Pulupa}, {Richter}, {Rodr{\'\i}guez-Pacheco}, {S{\'a}nchez-Cano},
  {Schuller}, {Ueno}, {Vainio}, {Vecchio}, {Veronig}, \&
  {Wijsen}}]{Dresing2023}
{Dresing}, N., {Rodr{\'\i}guez-Garc{\'\i}a}, L., {Jebaraj}, I.~C., {et~al.}
  2023, \aap, 674, A105

\bibitem[{{Echer}(2019)}]{Echer2019}
{Echer}, E. 2019, \grl, 46, 5681

\bibitem[{{Fox} {et~al.}(2016){Fox}, {Velli}, {Bale}, {Decker}, {Driesman},
  {Howard}, {Kasper}, {Kinnison}, {Kusterer}, {Lario}, {Lockwood}, {McComas},
  {Raouafi}, \& {Szabo}}]{fox16}
{Fox}, N.~J., {Velli}, M.~C., {Bale}, S.~D., {et~al.} 2016, \ssr, 204, 7

\bibitem[{Gieseler {et~al.}(2023)Gieseler, Dresing, Palmroos, Freiherr~von
  Forstner, Price, Vainio, Kouloumvakos, Rodríguez-García, Trotta, Génot,
  Masson, Roth, \& Veronig}]{giesler2023}
Gieseler, J., Dresing, N., Palmroos, C., {et~al.} 2023, Frontiers in Astronomy
  and Space Sciences, 9

\bibitem[{Gogosov(1961)}]{Gogosov61}
Gogosov, V. 1961, Journal of Applied Mathematics and Mechanics, 25, 148

\bibitem[{{G{\'o}mez-Herrero} {et~al.}(2015){G{\'o}mez-Herrero}, {Dresing},
  {Klassen}, {Heber}, {Lario}, {Agueda}, {Malandraki}, {Blanco},
  {Rodr{\'{\i}}guez-Pacheco}, \& {Banjac}}]{gomez-herrero15}
{G{\'o}mez-Herrero}, R., {Dresing}, N., {Klassen}, A., {et~al.} 2015, \apj,
  799, 55

\bibitem[{{Harvey} {et~al.}(1996){Harvey}, {Hill}, {Hubbard}, {Kennedy},
  {Leibacher}, {Pintar}, {Gilman}, {Noyes}, {Title}, {Toomre}, {Ulrich},
  {Bhatnagar}, {Kennewell}, {Marquette}, {Patron}, {Saa}, \&
  {Yasukawa}}]{harvey96}
{Harvey}, J.~W., {Hill}, F., {Hubbard}, R.~P., {et~al.} 1996, Science, 272,
  1284

\bibitem[{{Hickmann} {et~al.}(2015){Hickmann}, {Godinez}, {Henney}, \&
  {Arge}}]{hickmann2015}
{Hickmann}, K.~S., {Godinez}, H.~C., {Henney}, C.~J., \& {Arge}, C.~N. 2015,
  \solphys, 290, 1105

\bibitem[{{Hou} {et~al.}(2022){Hou}, {Tian}, {Wang}, {Zhang}, {Song}, {Zheng},
  {Chen}, {Chen}, {Bai}, {Chen}, {He}, {Song}, {Zhang}, {Hu}, {Dun}, {Zong},
  {Song}, {Xu}, \& {Tan}}]{Hou2022}
{Hou}, Z., {Tian}, H., {Wang}, J.-S., {et~al.} 2022, \apj, 928, 98

\bibitem[{{Howard} {et~al.}(2008){Howard}, {Moses}, {Vourlidas}, {Newmark},
  {Socker}, {Plunkett}, {Korendyke}, {Cook}, {Hurley}, {Davila}, {Thompson},
  {St Cyr}, {Mentzell}, {Mehalick}, {Lemen}, {Wuelser}, {Duncan}, {Tarbell},
  {Wolfson}, {Moore}, {Harrison}, {Waltham}, {Lang}, {Davis}, {Eyles},
  {Mapson-Menard}, {Simnett}, {Halain}, {Defise}, {Mazy}, {Rochus}, {Mercier},
  {Ravet}, {Delmotte}, {Auchere}, {Delaboudiniere}, {Bothmer}, {Deutsch},
  {Wang}, {Rich}, {Cooper}, {Stephens}, {Maahs}, {Baugh}, {McMullin}, \&
  {Carter}}]{howard2008}
{Howard}, R.~A., {Moses}, J.~D., {Vourlidas}, A., {et~al.} 2008, \ssr, 136, 67

\bibitem[{Hugoniot(1887)}]{hugoniot1887propagation}
Hugoniot, H. 1887, Journal de l'\'Ecole Polytechnique, CLVII, 3

\bibitem[{Hugoniot(1889)}]{hugoniot1889propagation}
Hugoniot, H. 1889, Journal de l'\'Ecole Polytechnique, CLVIII, 1

\bibitem[{{Jebaraj} {et~al.}(2024){Jebaraj}, {Agapitov}, {Krasnoselskikh},
  {Vuorinen}, {Gedalin}, {Choi}, {Palmerio}, {Wijsen}, {Dresing}, {Cohen},
  {Kouloumvakos}, {Balikhin}, {Vainio}, {Kilpua}, {Afanasiev}, {Verniero},
  {Mitchell}, {Trotta}, {Hill}, {Raouafi}, \& {Bale}}]{Jebaraj2024}
{Jebaraj}, I.~C., {Agapitov}, O., {Krasnoselskikh}, V., {et~al.} 2024, \apjl,
  968, L8

\bibitem[{{Jebaraj} {et~al.}(2021){Jebaraj}, {Kouloumvakos}, {Magdaleni{\'c}},
  {Rouillard}, {Mann}, {Krupar}, \& {Poedts}}]{Jebaraj21}
{Jebaraj}, I.~C., {Kouloumvakos}, A., {Magdaleni{\'c}}, J., {et~al.} 2021,
  \aap, 654

\bibitem[{{Kaiser} {et~al.}(2008){Kaiser}, {Kucera}, {Davila}, {St. Cyr},
  {Guhathakurta}, \& {Christian}}]{kaiser08}
{Kaiser}, M.~L., {Kucera}, T.~A., {Davila}, J.~M., {et~al.} 2008, \ssr, 136, 5

\bibitem[{{Kollhoff} {et~al.}(2021){Kollhoff}, {Kouloumvakos}, {Lario},
  {Dresing}, {G{\'o}mez-Herrero}, {Rodr{\'\i}guez-Garc{\'\i}a}, {Malandraki},
  {Richardson}, {Posner}, {Klein}, {Pacheco}, {Klassen}, {Heber}, {Cohen},
  {Laitinen}, {Cernuda}, {Dalla}, {Espinosa Lara}, {Vainio}, {K{\"o}berle},
  {K{\"u}hl}, {Xu}, {Berger}, {Eldrum}, {Br{\"u}dern}, {Laurenza}, {Kilpua},
  {Aran}, {Rouillard}, {Bu{\v{c}}{\'\i}k}, {Wijsen}, {Pomoell},
  {Wimmer-Schweingruber}, {Martin}, {B{\"o}ttcher}, {Freiherr von Forstner},
  {Terasa}, {Boden}, {Kulkarni}, {Ravanbakhsh}, {Yedla}, {Janitzek},
  {Rodr{\'\i}guez-Pacheco}, {Prieto Mateo}, {S{\'a}nchez Prieto}, {Parra
  Espada}, {Rodr{\'\i}guez Polo}, {Mart{\'\i}nez Hell{\'\i}n}, {Carcaboso},
  {Mason}, {Ho}, {Allen}, {Bruce Andrews}, {Schlemm}, {Seifert}, {Tyagi},
  {Lees}, {Hayes}, {Bale}, {Krupar}, {Horbury}, {Angelini}, {Evans}, {O'Brien},
  {Maksimovic}, {Khotyaintsev}, {Vecchio}, {Steinvall}, \&
  {Asvestari}}]{Kollhoff2021}
{Kollhoff}, A., {Kouloumvakos}, A., {Lario}, D., {et~al.} 2021, \aap, 656, A20

\bibitem[{{Kouloumvakos} {et~al.}(2022){Kouloumvakos}, {Kwon},
  {Rodr{\'\i}guez-Garc{\'\i}a}, {Lario}, {Dresing}, {Kilpua}, {Vainio},
  {T{\"o}r{\"o}k}, {Plotnikov}, {Rouillard}, {Downs}, {Linker}, {Malandraki},
  {Pinto}, {Riley}, \& {Allen}}]{Kouloumvakos2022}
{Kouloumvakos}, A., {Kwon}, R.~Y., {Rodr{\'\i}guez-Garc{\'\i}a}, L., {et~al.}
  2022, \aap, 660, A84

\bibitem[{{Kouloumvakos} {et~al.}(2019){Kouloumvakos}, {Rouillard}, {Wu},
  {Vainio}, {Vourlidas}, {Plotnikov}, {Afanasiev}, \&
  {{\"O}nel}}]{kouloumvakos19}
{Kouloumvakos}, A., {Rouillard}, A.~P., {Wu}, Y., {et~al.} 2019, \apj, 876, 80

\bibitem[{{Kwon} \& {Vourlidas}(2017)}]{Kwon2017}
{Kwon}, R.-Y. \& {Vourlidas}, A. 2017, \apj, 836, 246

\bibitem[{{Kwon} \& {Vourlidas}(2018)}]{Kwon2018}
{Kwon}, R.-Y. \& {Vourlidas}, A. 2018, Journal of Space Weather and Space
  Climate, 8, A08

\bibitem[{Landau \& Lifshitz(1959)}]{landau1959fluid}
Landau, L.~D. \& Lifshitz, E.~M. 1959, Course of Theoretical Physics, Vol.~6,
  Fluid Mechanics (Oxford: Pergamon Press)

\bibitem[{{Lario} {et~al.}(2017){Lario}, {Kwon}, {Riley}, \&
  {Raouafi}}]{lario17}
{Lario}, D., {Kwon}, R.-Y., {Riley}, P., \& {Raouafi}, N.~E. 2017, \apj, 847,
  103

\bibitem[{{Lario} {et~al.}(2016){Lario}, {Kwon}, {Vourlidas}, {Raouafi},
  {Haggerty}, {Ho}, {Anderson}, {Papaioannou}, {G\'omez-Herrero}, {Dresing}, \&
  {Riley}}]{lario16}
{Lario}, D., {Kwon}, R.-Y., {Vourlidas}, A., {et~al.} 2016, \apj, 819, 72

\bibitem[{{Lario} {et~al.}(2014){Lario}, {Raouafi}, {Kwon}, {Zhang},
  {G{\'o}mez-Herrero}, {Dresing}, \& {Riley}}]{Lario2014}
{Lario}, D., {Raouafi}, N.~E., {Kwon}, R.-Y., {et~al.} 2014, \apj, 797, 8

\bibitem[{{Li} {et~al.}(2024){Li}, {Zhao}, {Yan}, {Wu}, {Yang}, {Lv}, {Feng},
  {Ruan}, {Xiang}, \& {Liang}}]{Li2024ApJ}
{Li}, R., {Zhao}, X., {Yan}, J., {et~al.} 2024, \apj, 962, 178

\bibitem[{{Linan} {et~al.}(2025){Linan}, {Baratashvili}, {Lani}, {Schmieder},
  {Brchnelova}, {Guo}, \& {Poedts}}]{linan2025}
{Linan}, L., {Baratashvili}, T., {Lani}, A., {et~al.} 2025, \aap, 693, A229

\bibitem[{{Liu} {et~al.}(2017){Liu}, {Hu}, {Zhu}, {Luhmann}, \&
  {Vourlidas}}]{Liu17}
{Liu}, Y.~D., {Hu}, H., {Zhu}, B., {Luhmann}, J.~G., \& {Vourlidas}, A. 2017,
  \apj, 834, 158

\bibitem[{{Maharana} {et~al.}(2022){Maharana}, {Isavnin}, {Scolini}, {Wijsen},
  {Rodriguez}, {Mierla}, {Magdaleni{\'c}}, \& {Poedts}}]{maharana2022}
{Maharana}, A., {Isavnin}, A., {Scolini}, C., {et~al.} 2022, Advances in Space
  Research, 70, 1641

\bibitem[{{Malkov} \& {Drury}(2001)}]{malkov01}
{Malkov}, M.~A. \& {Drury}, L.~O. 2001, Reports on Progress in Physics, 64, 429

\bibitem[{{Manchester} {et~al.}(2004){Manchester}, {Gombosi}, {Roussev},
  {Ridley}, {de Zeeuw}, {Sokolov}, {Powell}, \&
  {T{\'o}th}}]{Manchester2004JGRA}
{Manchester}, W.~B., {Gombosi}, T.~I., {Roussev}, I., {et~al.} 2004, Journal of
  Geophysical Research (Space Physics), 109, A02107

\bibitem[{{Moon} {et~al.}(2002){Moon}, {Dryer}, {Smith}, {Park}, \&
  {Cho}}]{Moon2002}
{Moon}, Y.~J., {Dryer}, M., {Smith}, Z., {Park}, Y.~D., \& {Cho}, K.~S. 2002,
  \grl, 29, 1390

\bibitem[{{M{\"u}ller} {et~al.}(2020){M{\"u}ller}, {Zouganelis}, {St. Cyr},
  {Gilbert}, \& {Nieves-Chinchilla}}]{muller20}
{M{\"u}ller}, D., {Zouganelis}, I., {St. Cyr}, O.~C., {Gilbert}, H.~R., \&
  {Nieves-Chinchilla}, T. 2020, Nat. Astron., 4, 205

\bibitem[{{Nitta} {et~al.}(2013){Nitta}, {Schrijver}, {Title}, \&
  {Liu}}]{Nitta2013}
{Nitta}, N.~V., {Schrijver}, C.~J., {Title}, A.~M., \& {Liu}, W. 2013, \apj,
  776, 58

\bibitem[{{Odstrcil}(2003)}]{odstrcil03}
{Odstrcil}, D. 2003, Adv. Space Res., 32, 497

\bibitem[{{Pinter} \& {Dryer}(1990)}]{Pinter1990}
{Pinter}, S. \& {Dryer}, M. 1990, Bulletin of the Astronomical Institutes of
  Czechoslovakia, 41, 137

\bibitem[{{Poedts} {et~al.}(2020){Poedts}, {Lani}, {Scolini}, {Verbeke},
  {Wijsen}, {Lapenta}, {Laperre}, {Millas}, {Innocenti}, {Chan{\'e}},
  {Baratashvili}, {Samara}, {Van der Linden}, {Rodriguez}, {Vanlommel},
  {Vainio}, {Afanasiev}, {Kilpua}, {Pomoell}, {Sarkar}, {Aran}, {Sanahuja},
  {Paredes}, {Clarke}, {Thomson}, {Rouilard}, {Pinto}, {Marchaudon}, {Blelly},
  {Gorce}, {Plotnikov}, {Kouloumvakos}, {Heber}, {Herbst}, {Kochanov},
  {Raeder}, \& {Depauw}}]{Poedts2020}
{Poedts}, S., {Lani}, A., {Scolini}, C., {et~al.} 2020, Journal of Space
  Weather and Space Climate, 10, 57

\bibitem[{{Pomoell} \& {Poedts}(2018)}]{pomoell18}
{Pomoell}, J. \& {Poedts}, S. 2018, JSWSC, 8, A35

\bibitem[{Rankine(1870)}]{rankine1870thermodynamic}
Rankine, W. J.~M. 1870, Philos. Trans. R. Soc., 160, 277

\bibitem[{{Rodr{\'\i}guez-Garc{\'\i}a}
  {et~al.}(2021){Rodr{\'\i}guez-Garc{\'\i}a}, {G{\'o}mez-Herrero},
  {Zouganelis}, {Balmaceda}, {Nieves-Chinchilla}, {Dresing}, {Dumbovi{\'c}},
  {Nitta}, {Carcaboso}, {dos Santos}, {Jian}, {Mays}, {Williams}, \&
  {Rodr{\'\i}guez-Pacheco}}]{2021Rodriguez-Garcia}
{Rodr{\'\i}guez-Garc{\'\i}a}, L., {G{\'o}mez-Herrero}, R., {Zouganelis}, I.,
  {et~al.} 2021, \aap, 653, A137

\bibitem[{{Rodr{\'\i}guez-Garc{\'\i}a}
  {et~al.}(2022){Rodr{\'\i}guez-Garc{\'\i}a}, {Nieves-Chinchilla},
  {G{\'o}mez-Herrero}, {Zouganelis}, {Vourlidas}, {Balmaceda}, {Dumbovi{\'c}},
  {Jian}, {Mays}, {Carcaboso}, {dos Santos}, \&
  {Rodr{\'\i}guez-Pacheco}}]{2022Rodriguez-Garcia}
{Rodr{\'\i}guez-Garc{\'\i}a}, L., {Nieves-Chinchilla}, T., {G{\'o}mez-Herrero},
  R., {et~al.} 2022, \aap, 662, A45

\bibitem[{{Rouillard} {et~al.}(2012){Rouillard}, {Sheeley}, {Tylka},
  {Vourlidas}, {Ng}, {Rakowski}, {Cohen}, {Mewaldt}, {Mason}, \&
  {Reames}}]{Rouillard2012}
{Rouillard}, A.~P., {Sheeley}, N.~R., {Tylka}, A., {et~al.} 2012, \apj, 752, 44

\bibitem[{{Samara} {et~al.}(2024){Samara}, {Arge}, {Pinto}, {Magdaleni{\'c}},
  {Wijsen}, {Stevens}, {Rodriguez}, \& {Poedts}}]{Samara2024}
{Samara}, E., {Arge}, C.~N., {Pinto}, R.~F., {et~al.} 2024, \apj, 971, 83

\bibitem[{{Scolini} \& {Palmerio}(2024)}]{scolini24}
{Scolini}, C. \& {Palmerio}, E. 2024, Journal of Space Weather and Space
  Climate, 14, 13

\bibitem[{{Scolini} {et~al.}(2019){Scolini}, {Rodriguez}, {Mierla}, {Pomoell},
  \& {Poedts}}]{scolini19}
{Scolini}, C., {Rodriguez}, L., {Mierla}, M., {Pomoell}, J., \& {Poedts}, S.
  2019, \aap, 626, A122

\bibitem[{{Sedov}(1946)}]{Sedov1946}
{Sedov}, L.~I. 1946, Journal of Applied Mathematics and Mechanics, 10, 241

\bibitem[{{Smart} \& {Shea}(1985)}]{Smart1985}
{Smart}, D.~F. \& {Shea}, M.~A. 1985, \jgr, 90, 183

\bibitem[{{Strauss} {et~al.}(2017){Strauss}, {Dresing}, \&
  {Engelbrecht}}]{strauss17b}
{Strauss}, R.~D.~T., {Dresing}, N., \& {Engelbrecht}, N.~E. 2017, \apj, 837, 43

\bibitem[{{Taylor}(1950{\natexlab{a}})}]{taylor1950II}
{Taylor}, G. 1950{\natexlab{a}}, Proceedings of the Royal Society of London
  Series A, 201, 159

\bibitem[{{Taylor}(1950{\natexlab{b}})}]{taylor1950I}
{Taylor}, G. 1950{\natexlab{b}}, Proceedings of the Royal Society of London
  Series A, 201, 175

\bibitem[{{Thernisien}(2011)}]{Thernisien2011}
{Thernisien}, A. 2011, \apjs, 194, 33

\bibitem[{{Thernisien} {et~al.}(2006){Thernisien}, {Howard}, \&
  {Vourlidas}}]{Thernisien2006GCS}
{Thernisien}, A.~F.~R., {Howard}, R.~A., \& {Vourlidas}, A. 2006, \apj, 652,
  763

\bibitem[{{Verbeke} {et~al.}(2019){Verbeke}, {Pomoell}, \&
  {Poedts}}]{verbeke19}
{Verbeke}, C., {Pomoell}, J., \& {Poedts}, S. 2019, \aap, 627, A111

\bibitem[{von Neumann(1963)}]{vonNeumann1963}
von Neumann, J. 1963, in Collected Works, Volume 6 (New York: Pergamon),
  219--237

\bibitem[{{Warmuth}(2015)}]{Warmuth2015}
{Warmuth}, A. 2015, Living Reviews in Solar Physics, 12, 3

\bibitem[{{Wijsen} {et~al.}(2022){Wijsen}, {Aran}, {Scolini}, {Lario},
  {Afanasiev}, {Vainio}, {Sanahuja}, {Pomoell}, \& {Poedts}}]{wijsen22}
{Wijsen}, N., {Aran}, A., {Scolini}, C., {et~al.} 2022, \aap, 659, A187

\bibitem[{{Wijsen} {et~al.}(2023){Wijsen}, {Lario}, {S{\'a}nchez-Cano},
  {Jebaraj}, {Dresing}, {Richardson}, {Aran}, {Kouloumvakos}, {Ding},
  {Niemela}, {Palmerio}, {Carcaboso}, {Vainio}, {Afanasiev}, {Pinto},
  {Pacheco}, {Poedts}, \& {Heyner}}]{Wijsen2023}
{Wijsen}, N., {Lario}, D., {S{\'a}nchez-Cano}, B., {et~al.} 2023, \apj, 950,
  172

\bibitem[{{Yashiro} {et~al.}(2004){Yashiro}, {Gopalswamy}, {Michalek}, {St.
  Cyr}, {Plunkett}, {Rich}, \& {Howard}}]{Yashiro2004}
{Yashiro}, S., {Gopalswamy}, N., {Michalek}, G., {et~al.} 2004, Journal of
  Geophysical Research (Space Physics), 109, A07105

\end{thebibliography}
\begin{appendix}
    
\section{Modelling freely propagating shock waves}\label{app:EUHFORIA}

EUHFORIA's MHD module solves the complete set of ideal MHD equations, including the conservation of mass, momentum, energy, and the induction equation for the magnetic field. The solar wind in the heliosphere is modelled by evolving these equations through a structured mesh. At 0.1~au, the inner boundary, the solar wind velocity is radial and superfast, meaning that the solar wind speed exceeds the local fast magnetosonic speed everywhere. This ensures that all MHD characteristics are directed outwards.
Furthermore, CMEs can be introduced at the inner boundary as hydrodynamic plasma blobs --- for example, the cone CME or the spheroid model \citep{scolini24} --- or as flux-rope-like structures --- for example, the spheromak \citep{verbeke19} or the FRi3D \citep{maharana2022} descriptions. These CMEs are characterized by parameters such as speed, angular width, latitude, longitude, and magnetic field flux, among others.

Within the EUHFORIA model, CMEs are inserted into the simulation by assigning the CME’s bulk velocity and plasma properties at the 0.1~au boundary, rather than directly specifying the shock speed \( V_s \). This approach effectively introduces a discontinuity at the inner boundary when the CME enters the computational domain, representing a Riemann problem rather than an initial shock wave \citep{Gogosov61,Akhiezer75}.  
If the inserted CME's speed exceeds the local fast magnetosonic speed in the upstream reference frame, a shock wave forms within the computational domain. At the moment of formation, this shock has zero standoff distance from the CME and lacks thus both the shock sheath and the proper lateral extent that would have developed in the solar corona. The shock speed, which will always exceed the CME insertion speed, is determined by the local upstream solar wind conditions and the plasma properties of the CME.

In contrast, to insert a fully resolved fast-mode shock wave with a specific speed (e.g. estimated from white-light images), we start by assuming a shock speed and use the upstream plasma conditions from the EUHFORIA simulation at the inner boundary. The RH \citep[][]{rankine1870thermodynamic,hugoniot1887propagation, hugoniot1889propagation} jump conditions for MHD shocks are applied to directly determine the downstream conditions. Assuming a reference frame comoving with the shock, the RH jump conditions can be expressed as
\begin{align}
& \left[ \rho v_n \right] = 0 \,\, , \label{eq:RH1} \\
& \left[\left(\frac{1}{2}\rho v_n^2 + \frac{\gamma}{\gamma -1}p + \frac{B_n^2}{2\mu_0}\right) v_n - \frac{\vec{v}\cdot\vec{B}}{\mu_0}B_n \right] = 0 \,\, , \\
& \left[\rho v_n^2 + p + \frac{ B_t^2}{2\mu_0} \right] = 0 \,\, , \label{eq:rh_momentum}\\
& \left[\rho v_n \vec{v}_t - \frac{ B_n}{\mu_0}\vec{B}_t \right] = 0 \,\, , \\
& [B_n] = 0 \,\, , \\
& [v_n\vec{B}_t - B_n \vec{v}_t] = 0 \,\, , \label{eq:RH6}
\end{align}

\noindent where \( \rho \) is the plasma density, \( \mathbf{v} \) is the velocity vector, \( \mathbf{B} \) is the magnetic field vector, \( p \) is the pressure, \( \gamma \) is the adiabatic index, and \( \mu_0 \) is the permeability of free space. The square brackets \([ \cdot ]\) denote the jump in the respective quantities across the shock front. The normal vector \( \mathbf{n} \) is oriented perpendicular to the shock front, and the subscripts `$n$' and `$t$' refer to the components of the vectors parallel and perpendicular to the normal direction, respectively.

Solving Eqs.~\eqref{eq:RH1} to~\eqref{eq:RH6}  provides the downstream plasma conditions, which we prescribe at the inner boundary of the EUHFORIA simulation to model the shock wave with the desired speed. In this work, we focus solely on fast-mode shocks. Additionally, in MHD, the shock compression ratio \( r = \rho_{\rm down} / \rho_{\rm up} \) approaches \( ({\gamma + 1})/({\gamma - 1})\) for strong shocks, where the fast magnetosonic Mach number \( {M}_{\text{fms}} \rightarrow \infty \). For a typical monoatomic gas with \( \gamma = 5/3 \), the compression ratio \( r \) tends towards 4. However, EUHFORIA assumes \( \gamma = 1.5 \), resulting in a  maximal compression ratio of \( r \rightarrow 5 \). Furthermore, because the magnetic field is almost purely radial at the inner boundary, the shock is what is known as a `parallel shock'. In such a configuration and within the assumptions of ideal MHD, the magnetic field has minimal influence on the shock's behaviour.

To return to the background solar wind state after the shock has passed in the EUHFORIA simulation, we applied the following decay function:
\begin{equation}
 y = y_s \exp\left[{\ln\left|\frac{y_{b}}{y_s}\right|\left(\frac{t}{\Delta t}\right)^\alpha}\right]\quad \text{with} \quad 0\leq t \leq \Delta t, \label{eq:decay}   
\end{equation}
where \(t\) denotes the time, \( y \) represents any of the plasma variables, \( y_s \) denotes the shocked state, and \( y_b \) represents the background solar wind state. The parameter \( \Delta t \) defines the time interval over which the plasma variable relaxes to its background state, which we chose in this work as $\Delta t = 20.5 R_\odot / V_s$, that is, the time the shock would need to propagate from the solar surface to the inner boundary of EUHFORIA (at $21.5 R_\odot$) assuming a constant speed.
The parameter \( \alpha \) in Eq.~\eqref{eq:decay} is a free parameter that controls the decay rate — smaller values of \( \alpha \) correspond to faster decays. This decay function is an ad hoc approximation, since a full solution to the downstream profile would require solving the MHD equations from the low corona onwards. Nevertheless, the exponential decay ensures that most of the shocked plasma remains concentrated directly downstream of the shock, qualitatively resembling the SVT blast wave solutions. Additionally, this formulation guarantees a smooth transition from the shock-affected region to the ambient solar wind conditions.

\section{Modified Sedov--von Neumann--Taylor solution}\label{app:sedov}

\begin{figure*}
\centering
    \includegraphics[width=0.85\linewidth]{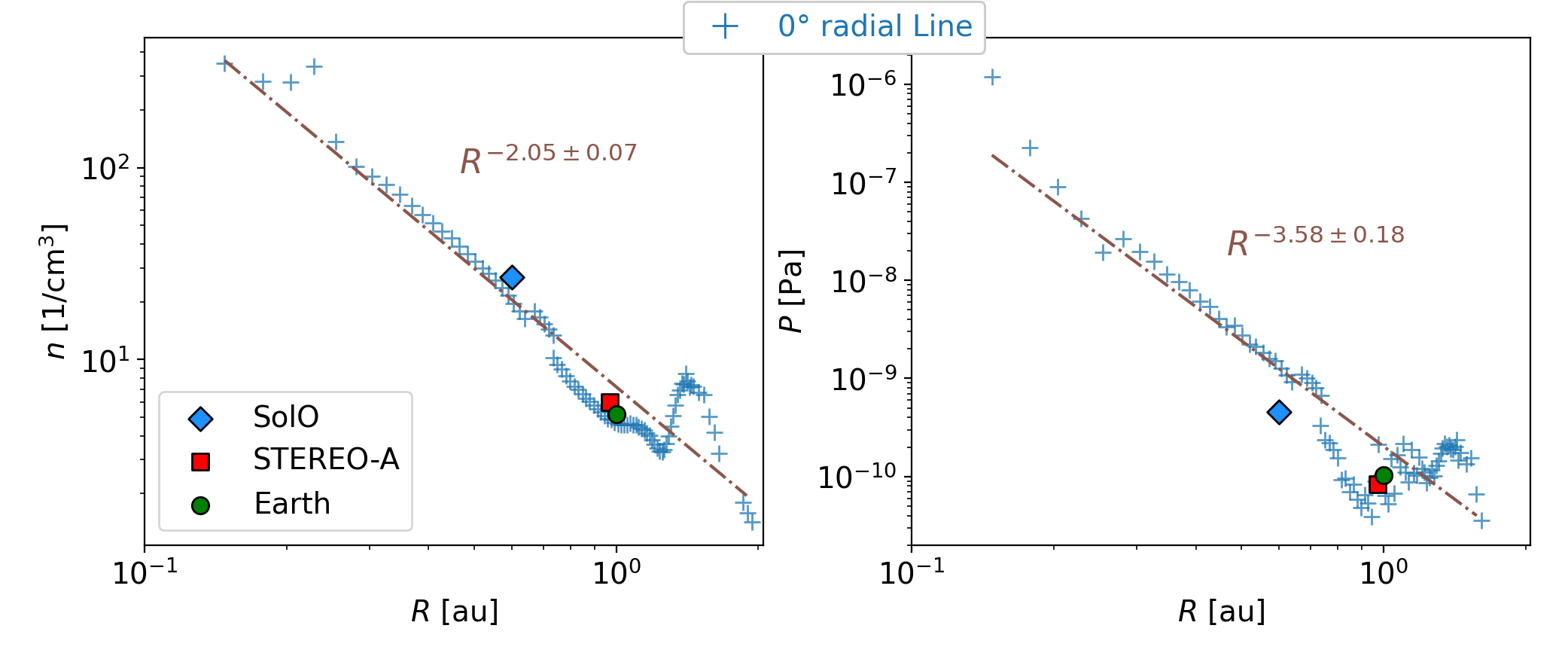}
    \caption{
Left panel: Upstream number density (\( n \)) as a function of shock location (\( R \)) along the $0^\circ$ radial line, with simulation data fitted by \( n \propto R^{-2.05 \pm 0.07} \). 
Right panel: Downstream pressure immediately behind the shock as a function of \( R \), with simulation data fitted by \( P \propto R^{-3.48 \pm 0.18} \). The symbols indicate in situ spacecraft data.
    }
    \label{fig:shock_scaling}
\end{figure*}

The SVT solution provides a self-similar description of the propagation of a strong shock wave resulting from an instantaneous, point-like energy release (e.g. explosions) into a homogeneous medium. Due to their simplicity, these solutions have been widely used in astrophysics, notably for describing supernova remnant shock waves. While the interstellar or terrestrial medium where explosions can occur is often assumed to be homogeneous, many astrophysical environments, including the solar wind, exhibit strong density gradients \citep{Cavaliere1976, Chevalier82}.

Once the shock decouples from its CME driver, its evolution becomes similar to a Sedov–Taylor blast wave. At this stage, the shock effectively loses memory of how it originally gained its energy — whether through a sustained piston-driving process or a sudden energy release. Instead, its subsequent evolution is governed by its own momentum and interaction with the ambient solar wind, with its deceleration described well by the Sedov–Taylor framework \citep[e.g.][]{Pinter1990,Moon2002,Li2024ApJ}.

Here, we derive a self-similar solution for a medium where the ambient density decreases with radius as \( \rho(r) = {Ar^{-\beta}} \), with \(A\) a proportionality constant. For the expanding solar wind, we can assume \(\beta = 2\) as a first approximation. To derive the self-similar solution, we assume spherical symmetry, a strong shock with a Mach number \(\mathcal{M} \gg 1\),  and energy conservation where the total energy \( E \) remains constant over time.

To determine the scaling law for the shock radius \( R(t) \) as a function of time \( t \), energy \( E \), and the ambient density parameter \( A \), we used a dimensional analysis similar to that of \cite{Sedov1946}. We began by assuming that \( R(t) \) has a power-law dependence on time:

\begin{equation}
    R(t) \propto t^{\alpha}, \label{eq:scaling_law}
\end{equation}
where our objective is to determine the exponent \( \alpha \).  

The total energy (\( E \)) can be expressed in terms of the post-shock quantities. For dimensional consistency, we have

\begin{equation}
    E \propto \rho(R) R^3 \left(\frac{dR}{dt}\right)^2. \label{eq:energy_scaling}
\end{equation}Substituting \( \rho(R) = \frac{A}{R^\beta} \) and \( R \propto t^{\alpha} \) into Eq.~\eqref{eq:energy_scaling}, we obtain

\begin{equation}
    E \propto A t^{(5 - \beta)\alpha - 2}. \label{eq:energy_const}
\end{equation}Since the total energy \( E \) is constant, we solve for \( \alpha \) as follows:

\begin{equation}
    \alpha = \frac{2}{5 - \beta}. \label{eq:alpha_value}
\end{equation}The shock speed (\( V(t) \)) is then given by

\begin{equation}
    V_\text{sh}(t) = \frac{dR}{dt} \propto t^{-({3 - \beta})/({5 - \beta})}. \label{eq:shock_velocity_gen}
\end{equation}The pressure immediately downstream of the shock can be estimated using the relation

\begin{equation}
    P \propto \rho V^2, \label{eq:pressure_relation}
\end{equation}which follows from the RH momentum jump condition \eqref{eq:rh_momentum}, assuming negligible upstream magnetic field and upstream pressure \citep{landau1959fluid}. Applying the scaling laws for \( V \) and \( \rho \), we find

\begin{equation}
    P(t) \propto t^{-6/(5-\beta)}. \label{eq:P_t_expression_gen}
\end{equation}In the specific case where \( \beta = 2 \), we have

\begin{equation}
    R(t) \propto t^{2/3}, \label{eq:shock_radius}
\end{equation}and thus the shock velocity,

\begin{equation}
    V_\text{sh}(t) = \frac{dR}{dt} \propto t^{-1/3}. \label{eq:shock_velocity}
\end{equation}

\noindent Accordingly, the downstream pressure evolves as

\begin{equation}
    P(t) \propto t^{-2}. \label{eq:pressure_time}
\end{equation}Expressing these in terms of the shock radius \( R \), using \( t \propto R^{1/\alpha} \), we obtain

\begin{equation}
    V_\text{sh}(R) \propto R^{-1/2}, \label{eq:shock_velocity_radius}
\end{equation}and

\begin{equation}
    P(R) \propto R^{-3}. \label{eq:pressure_radius}
\end{equation}

In summary, we find that the shock radius expands as \( t^{2/3} \), indicating a decelerating expansion. This leads to a reduction in shock velocity over time as \( t^{-1/3} \) or with distance as \( R^{-1/2} \), consistent with, for example, \cite{Cavaliere1976} and \cite{Smart1985}. Meanwhile, the downstream pressure diminishes rapidly, scaling as \( t^{-2} \) or \( R^{-3} \).

These theoretical predictions align well with the observed trends shown in Fig.~\ref{fig:shock_scaling}, where the fitted power laws are \( n \propto R^{-2.05 \pm 0.07} \) for the upstream density and \( P \propto R^{-3.58 \pm 0.18} \) for the downstream pressure. While the power laws capture the average decay rates approximately, significant deviations are introduced by the preceding CMEs, which create fluctuations around these mean trends.

In addition, in situ measurements from various spacecraft are included in Fig.~\ref{fig:shock_scaling}. It is important to note, however, that none of the spacecraft are exactly positioned along the $0^\circ$ line in the solar equatorial plane. Earth, for instance, is closest to this line but is situated approximately 7 degrees south of the equatorial plane. Despite these positional discrepancies, the in situ data generally exhibit a similar decreasing trend, particularly in density.

For the downstream thermal pressure, both the simulation and observations display a comparable decaying trend; however, the simulations consistently yield higher pressure values than those observed. This difference is also evident in Fig.~\ref{fig:euh obs}, which compares the simulated and observed pressure time profiles. The likely causes for this discrepancy include a slight overestimation of the shock strength in the simulations, as well as an underestimation of the magnetic field magnitude, leading to a lower magnetic pressure in the simulation compared to actual observations.

\end{appendix}

\end{document}